\pdfoutput=1
\documentclass[a4paper,12pt]{article}
\usepackage[utf8]{inputenc}
\usepackage{amssymb,amsmath,amsfonts}
\usepackage{placeins}
\usepackage{xcolor}
\usepackage{dsfont}
\usepackage{graphicx}
\usepackage{hyphenat}
\usepackage{wrapfig}
\usepackage{empheq}
\usepackage{float}
\usepackage{setspace}
\usepackage{subcaption}
\usepackage{rotating}
\usepackage{pdfpages}
\usepackage{verbatim}
\usepackage{array}
\usepackage{caption}
\usepackage{braket}
\usepackage{cite}
\usepackage{epsfig}
\usepackage[left=2.4cm,top=3.3cm,right=2.4cm,bottom=3.3cm,bindingoffset=0cm]{geometry}
\usepackage[titletoc,page]{appendix}
\usepackage[colorlinks,linkcolor=purple,citecolor=teal]{hyperref}

\newcommand{\beq}{\begin{eqnarray}}
\newcommand{\eeq}{\end{eqnarray}}
\newcommand{\bea}{\begin{eqnarray}}
\newcommand{\eea}{\end{eqnarray}}
\newcommand{\be}{\begin{equation}}
\newcommand{\ee}{\end{equation}}

\renewcommand{\i}{\mathrm{i}}
\newcommand{\br}{\mathbf{r}}
\newcommand{\bu}{\mathbf{u}}

\def\de{\partial}

\def\1{\mathbbm{1}}

\def\c{\rm c}

\def\S{\mathbb{S}}

\def\nn{\nonumber}

\def\Tr{\qopname\relax o{Tr}}
\def\tr{\qopname\relax o{tr}}
\def\Vol{\qopname\relax o{Vol}}
\def\KK{\qopname\relax o{KK}}
\def\YM{\qopname\relax o{YM}}
\def\CS{\qopname\relax o{CS}}

\def\Nc{{N_{\rm c}}}
\def\Nf{{N_{\rm f}}}

\def\non{\nonumber\\}
\renewcommand{\d}{\mathrm{d}}
\def\Li{\mathrm{Li}}
\def\calG{\mathcal{G}}
\def\calA{\mathcal{A}}
\def\calB{\mathcal{B}}

\numberwithin{equation}{section}

\newtheorem{lemma}{Lemma}
\newtheorem{observation}{Observation}
\newtheorem{corollary}{Corollary}

\newcommand{\hA}[0]{\widehat{A}}
\newcommand{\hF}[0]{\widehat{F}}

\newcommand{\mA}[0]{\mathcal{A}}	
\newcommand{\mU}[0]{\mathcal{U}}

\begin{document}
\title{
\hfill\small {IFUP-TH-2021}\\[10pt]
\bf{\Large Holographic Nuclear Physics with Massive Quarks}}
\author{
  Salvatore Baldino$^{(1)}$,
  Lorenzo Bartolini$^{(2)}$,\\[0pt]
  Stefano Bolognesi$^{(3)}$ and
  Sven Bjarke Gudnason$^{(4)}$\\[20pt]
{\em \normalsize
$^{(1)}$Center for Mathematical Analysis, Geometry and Dynamical
  Systems,}\\[0pt]
{\em \normalsize
  Department of Mathematics, Instituto Superior T\'ecnico,}\\[0pt]
{\em \normalsize
  Universidade de Lisboa, Av. Rovisco Pais, 1049-001 Lisboa, Portugal
}\\
{\em \normalsize
$^{(2)}$ Institut f\"ur Theoretische Physik, Technische Universit\"at Wien,}\\[0pt]
{\em \normalsize
  Wiedner Hauptstrasse 8-10, A-1040 Vienna, Austria
}\\
{\em \normalsize
$^{(3)}$Department of Physics ``E. Fermi'', University of Pisa
  and INFN Sezione di Pisa}\\[0pt]
{\em \normalsize
  Largo Pontecorvo, 3, Ed. C, 56127 Pisa, Italy}\\[3pt]
{\em \normalsize
$^{(4)}$Institute of Contemporary Mathematics, School of
  Mathematics and Statistics,}\\[0pt]
{\em \normalsize Henan University, Kaifeng, Henan 475004,
  P.~R.~China}\\[10pt]
{\normalsize emails: salvatore.baldino(at)tecnico.ulisboa.pt, lorenzobartolini89(at)gmail.com,  } \\
{ \normalsize stefano.bolognesi(at)unipi.it, gudnason(at)henu.edu.cn} \\
{\normalsize
}}
\date{June 2021}
\maketitle
\thispagestyle{empty}
\setcounter{page}{0}
\begin{abstract}
We discuss nuclear physics in the Witten-Sakai-Sugimoto model, in the
limit of large number $N_c$ of colors and large 't Hooft coupling,
with the addition of a finite mass for the quarks.
Individual baryons are described by classical solitons whose size is
much smaller than the typical distance in nuclear bound states, thus we
can use the linear approximation to compute the interaction potential
and provide a natural description for lightly bound states.   
We find the classical geometry of nuclear bound states for baryon
numbers up to $B=8$.  
The effect of the finite pion mass -- induced by the quark mass via
the GMOR relation -- is to decrease the binding energy of the nuclei
with respect to the massless case. 
We discuss the finite density case with a particular choice of a cubic
lattice, for which we find the critical chemical potential, at which
the hadronic phase transition occurs. 
\end{abstract}

\newpage

\tableofcontents

\section{Introduction}

The holographic model of Witten-Sakai-Sugimoto (WSS)
\cite{Witten:1998zw,Sakai:2004cn,Sakai:2005yt} is the top-down
holographic theory closest to QCD to date. 
The model is based on a D4--D8 brane setup in type IIA string theory
and the flavor dynamics is encoded in the low-energy action for the
gauge fields on the D8 flavor branes in the geometry left by the gauge 
D4-branes.
The model at low-energy reproduces a large-$\Nc$ SU($\Nc$) gauge
theory with $\Nf$ massless quarks plus a tower of massive adjoint
matter fields. 
Being a top-down model, it has very few parameters: $\Nc$, $\Nf$, the 
't Hooft coupling $\lambda$ and the mass scale $M_{\KK}$.
In this paper, we will also include a quark mass term. 
The model shares all the important features with QCD, in particular
confinement and chiral symmetry breaking as well as the existence of a 
low-energy chiral Lagrangian, which is of a Skyrme-type theory coupled
to an infinite tower of vector mesons. 
Baryons in the WSS model are identified with instantons of the gauge
theory describing the flavor branes
\cite{Hong:2007kx,Hata:2007mb,Hashimoto:2008zw,Bolognesi:2013nja},
just like baryons can be seen as solitons of the Skyrme model
\cite{Adkins:1983ya,Manko:2007pr}. 
Quantization of low-energy instantonic degrees of freedom provides
quantum numbers for the corresponding nucleons.
Many techniques developed in the context of quantization of zeromodes
of Skyrmions as nuclei have been used in our approach, see
e.g.~Refs.~\cite{Adkins:1983nw,Sutcliffe:2008sk,Jackson:1985bn,Braaten:1988cc,Verbaarschot:1986qi,Piette:1994ug,Leese:1994hb,Irwin:1998bs}.

Composite nuclei are described by multi-instantons configurations. 
We used this approach to describe composite nuclei from a
``solitonic perspective'' in a previous work \cite{Baldino:2017mqq},
where we restricted to the case of massless quarks.
This approach can be viewed as complementary to other approaches to
holographic nuclear physics, see
e.g.~Refs.~\cite{Hashimoto:2009ys,Kim:2009sr,Kim:2008iy,Kim:2011ey,Pahlavani:2010zzb,Pahlavani:2014dma,Hashimoto:2010je,Bolognesi:2013jba}.
In the limit which we are considering, the instanton radius scales as 
$\lambda^{-\frac{1}{2}}$ and the distances between individual nuclei
in the bound state configuration scale as $\lambda^0$ and so a linear
approach can be used for the computation of the dominant two-body
potential between the nuclei as an infinite sum of monopole and
dipole interactions.
In this limit the instantons become point-like but with an SU(2)
orientation, which is very similar to the models considered in 3+1
dimensions in Refs.~\cite{Gillard:2015eia,Gillard:2016esy}. 
In this work we include the quark masses, which via the
Gell-Man--Oakes--Renner (GMOR) relation induces a pion mass, which in
turn is felt by the solitonic (pionic) degrees of freedom. 
In Ref.~\cite{Baldino:2017mqq}, we found bound states in the large
$N_c$ and large $\lambda$ limit and computed their respective nuclear
binding energies.
In distinction to the previous work (i.e.~the massless case), the
inclusion of a pion mass makes the nuclear bound states larger and the
corresponding binding energies smaller.

The paper is organized as follows.
In section \ref{due} we give an overview of the model and discuss the
introduction of the quarks mass term in the framework.
In section \ref{quattro} we compute the nucleon-nucleon potential. 
In section \ref{cinquebis} we present the numerical results for the
nuclear bound states. 
In section \ref{sei} we discuss the quantization for the deuteron
state. 
In section \ref{sette} we discuss the case of finite density and the
hadronic phase transition. 
We conclude in section \ref{conclusion} with a discussion and outlook.

\section{The WSS holographic model with massive quarks}
\label{due}

The model encodes color degrees of freedom in a background metric from
type IIA string theory \cite{Witten:1998zw}: 
\begin{align}\label{metric}
ds^2 &= \left(\frac{u}{R}\right)^{\frac{3}{2}} \left(\eta_{\mu\nu}dx^{\mu}dx^{\nu} + f(u)dx_4^2\right)
+\left(\frac{R}{u}\right)^{\frac{3}{2}}\left( \frac{du^2}{f(u)} + u^2 d\Omega_4^2 \right) \;, \nn \\
e^{\phi} &= g_{\rm s} \left( \frac{u}{R} \right)^{\frac{3}{4}}\;, \quad\quad  F_4=dC_3= \frac{2\pi\Nc}{\Vol_4} \epsilon_4\;, \quad\quad
f(u) = 1 - \frac{u_{\KK}^3}{u^3}\;,
\end{align}
where $\phi$ is the dilaton, $C_n$, $F_{n+1}$ indicate the
Ramond-Ramond $n$-form and its corresponding field strength, 
${\rm Vol}_4$ and $\epsilon_4$ stand for the volume of a unit
radius 4-sphere ($S^4$) and its volume form, respectively. 
The function $f(u)$ makes the geometry terminate at a fixed coordinate 
$u_{\KK}$, so in order to avoid singularities, the coordinate $x_4$
has to be periodic with period 
\bea
\delta x_4 = \frac{4\pi}{3}\frac{R^{\frac{3}{2}}}{u_{\KK}^{\frac{1}{2}}} \equiv \frac{2\pi}{M_{\KK}}\;.
\eea
Throughout this article we will work in units of the intrinsic mass scale of the theory, that is, we set:
\beq
M_{\KK}= u_{\KK}=1\;.
\eeq

The inclusion of flavor degrees of freedom is performed via insertion
of a couple of $N_f$ stacked D8/$\overline{\rm D8}$-branes 
(with $N_f$ being the number of light quark flavors), transverse to
the color branes in the $x_4$ direction (that is, localized on the
circle) \cite{Sakai:2004cn,Sakai:2005yt}.
We will work in the setup with antipodal flavor branes, in which case
the two stacks merge at the cigar tip labeled by $u=u_{\KK}$: this is
in every sense a geometrical realization of the spontaneous breaking
of chiral symmetry. 
A rigorous treatment should include also the backreaction on the
geometry due to the presence of these stacks of branes, but since we
will only be considering the presence of two light flavors, $N_f=2$,
the effect of the modified geometry can be neglected as a first
approximation (see Ref.~\cite{Bigazzi:2014qsa} for a treatment of the 
backreaction of the flavor branes and the implications).

We will be interested in the theory on the D8/$\overline{\rm D8}$-branes,
so we employ for the cigar subspace bulk coordinates, defined by 
\begin{equation}
\left\{
\begin{aligned}
\quad  & u^3=u_{\KK}^3 + u_{\KK}r^2\;, \\
\quad & x_4= \frac{2 R^{\frac{3}{2}}}{3 u_{\KK}^{\frac{1}{2}}} \theta\;,
\end{aligned} \right.
\quad\Rightarrow \quad\left\{
\begin{aligned}
\quad  & y=r \cos \theta,\\
\quad & z=r \sin \theta,
\end{aligned}
\right.
\end{equation}
with $z$ running on the curve that defines the embedding of the flavor
branes, and $y$ being its transverse coordinate.
The action on the D8-brane world volume is composed by two terms: a
Yang-Mills action in warped spacetime arising from the truncation of
the DBI action to quadratic terms, and a Chern-Simons 
term, originating from the coupling of the D8-branes to the
Ramond-Ramond 3-form $C_3$:
\begin{align}
\label{Sefftot}
S&= S_{\YM} + S_{\CS}\;,\nonumber \\
S_{\YM} &=  -\kappa \Tr\int d^4 x\, dz \left[ \frac{1}{2}h(z)\mathcal{F}_{\mu \nu}^2 + k(z) \mathcal{F}_{\mu z}^2 \right]\;, \nn \\
S_{\CS} &=  \frac{\Nc}{384 \pi^2} \epsilon^{\alpha_1 \alpha_2 \alpha_3 \alpha_4 \alpha_5} \int d^4 x\, dz\; \hA_{\alpha_1} \left[ 6\tr\left(F_{\alpha_2 \alpha_3}^a F_{\alpha_4 \alpha_5}^a \right)+2\tr\big(\hF_{\alpha_2 \alpha_3}\hF_{\alpha_4 \alpha_5}\big)\right]\;,
\end{align}
with the warp factors $k(z),h(z)$ and $\kappa$ given by

\begin{equation}
k(z) = 1+z^2,\qquad h(z) = (1+z^2)^{-\frac{1}{3}},\qquad \kappa \equiv\frac{N_c \lambda}{216 \pi^3}\;.
\end{equation}
The notation for the indices we use is as follows:
\begin{align}
\alpha_1, \alpha_2,... &= \left\{0,1,2,3,z\right\}, &\qquad M,N,... &= \left\{1,2,3,z\right\}\;, \nonumber \\
i,j,k,... &= \left\{1,2,3\right\},&\qquad \mu,\nu,... &= \left\{0,1,2,3\right\}\;.
\end{align}

In continuity with Ref.~\cite{Baldino:2017mqq}, it is useful to define
a new coupling $\Lambda$, and a unique warp factor $H(z)$ as: 
\beq\label{LAMBDA}
\Lambda= \frac{8 \lambda}{27\pi}\;, \qquad
H(z) = \left(1+z^2\right)^{\frac{2}{3}}\;.
\eeq
Moreover, we rescale the action by:
\beq\label{rescale}
\S = \kappa^{-1} S\;,
\eeq
so that the full rescaled action in the new notation reads:
\begin{align}
\S &= -\frac{1}{2}\int d^4x\, dz\; H^{\frac{3}{2}} \left(\frac{1}{2}\hF_{\alpha_1 \alpha_2}\hF^{\alpha_1\alpha_2} + \tr\left(F_{\alpha_1\alpha_2}F^{\alpha_1\alpha_2}\right)\right)\nn \\
&\phantom{=\ }
\mathop+\frac{1}{\Lambda}\epsilon^{\alpha_1\alpha_2\alpha_3\alpha_4\alpha_5}\int d^4x\, dz\; \hA_{\alpha_1} \left(F_{\alpha_2\alpha_3}F_{\alpha_4\alpha_5} + \frac{1}{6}\hF_{\alpha_2\alpha_3} \hF_{\alpha_4\alpha_5}\right)\;.
\end{align}

\subsection{Quark mass}
\label{tre}

The addition of a quark mass to the model is performed by the insertion
of a Wilson line operator in the dual QCD-like theory: this is the
best we can do given the geometry, since flavor and anti-flavor
degrees of freedom are not localizable at the same position (they will
always remain separated along the $x_4$ direction). Hence, a quark mass
term will be nonlocal and take the form: 
\begin{equation}\label{deltaLmass}
\delta S \propto \int d^4x\; \sum_{i=1}^{N_f} OW_i^i (x)+{\rm h.c.}\;,
\end{equation}
where $OW_i^j$ is the open Wilson line operator:
\begin{equation}\label{OWL}
OW^j_i \equiv\psi_{\rm L}^{\dag j}\Big(x^{\mu}, x^4=-\frac{\pi}{2}R\Big) \,\mathcal{P}\exp\left[\int dx^4\; (\i\mathcal{A}_4 + \Phi)\right] \psi_{i\rm R} \Big(x^{\mu}, x^4= +\frac{\pi}{2}R\Big)\;.
\end{equation}
To obtain this object, we insert an open string stretching between the
flavor branes, and provide the action with the Aharony-Kutasov term
\cite{Aharony:2008an}:
\begin{equation}
S_{\rm AK}= \frac{\kappa'}{{\rm Vol}_4} \int d^4x\,d^4\Omega\; \sum_i e^{-S_{\rm str}^i}+{\rm h.c.}\;,
\end{equation}
with $S_{\rm str}$ being the action of the stretched string. The action
$S_{\rm str}$ is composed of two terms: the Nambu-Goto action for the
free string ($S_{\rm NG}$) and an interaction term of the string
endpoints with the flavor branes to which they are attached:
\beq
\frac{N'}{(2\pi)^3 g_s l_s^4}\left(\frac{u_{\rm KK}}{R}\right)^{\frac{9}{4}}\int d^4x\; e^{-S_{\rm NG}}\tr\left[\left(\mathcal{P}e^{-\i\int_{\de\rm WS} \mathcal{A}}-\mathds{1}\right)+{\rm h.c.}\right]\;,
\eeq
where the subtraction of the identity matrix accounts for the
subtraction of the vacuum.  
After factorizing away the contribution from the Nambu-Goto term in
$S_{\rm str}$, we are left with an action that includes the
interaction of the endpoint of the string with the $D8$ world volume:
the Nambu-Goto part will produce the quark mass and chiral condensate,
as well as prefactors. We pack all this information into the constant
$c$ and in the quark mass matrix $M$. Finally, since the 
endpoints are forced to move on the $y=0$ curve in the cigar subspace,
the action in terms of the gauge fields is:  
\begin{align}
S_{\rm AK} &= c\int d^4x\; \tr \mathcal{P}\left[\left(Me^{-\i\int_{-\infty}^{+\infty} dz \mathcal{A}_z} -\mathds{1}\right) + {\rm h.c.}\right], \qquad
c= \frac{\lambda^{\frac{3}{2}}}{3^{\frac{9}{2}}\pi^{3}}\;. \label{SAK}
\end{align}
At this point it is useful to evaluate $c$ in terms of $\Lambda$:
\beq
c =  \frac{\Lambda^{\frac{3}{2}}}{16\sqrt{2}\pi^{\frac{3}{2}}}\;,
\eeq
and to perform the rescaling (\ref{rescale}):
\beq
\S_{\rm AK}=\frac{2\left(2\Lambda\pi\right)^{\frac{1}{2}}}{N_c} \int d^4x\; \tr \mathcal{P}\left[M\left(e^{-\i\int_{-\infty}^{+\infty} dz \mathcal{A}_z} + e^{\i\int_{-\infty}^{+\infty} dz \mathcal{A}_z} -2\mathds{1} \right)\right]\;.
\eeq

\subsection{Deformation of the instanton}
\label{deformation}

The additional term in the Lagrangian induces two effects: it obviously
changes the mass of the 1-instanton configuration, but could also
potentially modify the size of the instanton. The classical YM
instanton possesses a modulus $\rho$, that is interpreted as the size
of the instanton and does not affect the energy. As explored in
Ref.~\cite{Hata:2007mb}, the combined effect of the Chern-Simons term
(that tends to dilate the instanton) and of the gravitational field
(that tends to shrink it) is reflected by the fact that $\rho$
ceases to be a modulus, and the energy gains a $\rho$ dependence. The
size of the instanton is then determined by minimizing the energy with
respect to $\rho$. Here we will determine the contribution of the mass
term to the total mass, and size of the instanton. We will follow the
computations made in Ref.~\cite{Hashimoto:2009hj}, adapting them to
the gauge that was used in Ref.~\cite{Bolognesi:2013nja}. 

The energy of the static model is obtained by computing the opposite
of the static action. This action is obtained by considering $A_I$ and
$\hat A_0$ as the only nonvanishing fields. The energy then reads
\begin{align}
\mathbb E=&\int d^3x\,dz\;\left(\tr\left( H(z)^{-\frac12}\frac12F_{ij}^2+H(z)^{\frac 32}F_{zi}^2\right)-\frac12H(z)^{-\frac12}(\partial_i \hat A_0)^2-\frac12 H(z)^{\frac32}(\partial_z\hat A_0)^2\right)\nonumber\\
&-\frac1\Lambda\int d^3x\,dz\;\hat A_0\tr(F_{IJ}F_{KL})\epsilon_{IJKL} \nonumber\\&-\frac{2\left(2\Lambda\pi\right)^{\frac{1}{2}}}{N_c} \int d^4x\, \tr \mathcal{P}\left[M\left(e^{-\i\int_{-\infty}^{+\infty} dz A_z} + e^{\i\int_{-\infty}^{+\infty} dz {A}_z}  -2\mathds{1}\right)\right]\;.
\label{totalene}
\end{align}
We will use the same field configuration as was used in
Ref.~\cite{Bolognesi:2013nja}. In fact, as argued in
Ref.~\cite{Hashimoto:2009hj}, modifications to the field profile are
subleading in $\Lambda$, and the first-order effects only affect the
energy.\footnote{
The implication of this is that the pions
receive a mass term, but the vector mesons are unaffected to leading
order in $1/\Lambda$.
}
The field configuration we use is  
\begin{align}
A_I=-\sigma_{IJ}x_Jb(\rho)\;,\qquad \hat A_0=a(\rho)\;,
\end{align}
where $\rho=\sqrt{x_Ix_I}$ and the profiles $a$ and $b$ are given by
\begin{align}
  a(r)=\frac8\Lambda\frac{r^2+2\rho^2}{(r^2+\rho^2)^2}\;,\qquad
  b(r)=\frac{1}{r^2+\rho^2}\;,
\end{align}
where $r=\sqrt{x_ix_i}$. 
With this configuration, we can perform the integrals. The integral of
the first two lines of Eq.~\eqref{totalene} is done in great detail in
Ref.~\cite{Bolognesi:2013nja}: resulting in: 
\begin{align}
8\pi^2\left(1+\frac{\rho^2}6+\frac{64}{5\Lambda^2\rho^2}\right) .
\end{align}
For the third line of Eq.~\eqref{totalene}, using the arguments of
Ref.~\cite{Hashimoto:2009hj} to get the leading order contribution in
$\Lambda$, we can write the integral as 
\begin{align}
\int d^3x\; \tr\big(M(U+U^\dagger-2\mathds 1)\big)\;.
\end{align}
To evaluate this integral in a divergenceless fashion, we must perform
a gauge transformation to avoid singularities: we will thus work in
the gauge 
\begin{align}
A_z=\left(\frac{1}{r^2+\rho^2}-\frac1{\rho^2}\right)x_i\sigma_i\;.
\end{align}
Performing the integration, the matrix exponentiation and the trace, we get
\begin{align}
 \tr\big(M(U+U^\dagger-2\mathds 1)\big)=-4m\left(1+\cos\pi\sqrt{\frac{r^2}{r^2+\rho^2}}\right)\;,
\end{align}
where we assumed the quark masses to be degenerate, and thus
$M=m\mathds{1}$, with $m_u=m_d=m$. 
This term cannot be integrated in closed form. The integral can be written as
\begin{align}
\int d^3x\;\tr\big(M(U+U^\dagger-2\mathds 1)\big)=-16m\pi\rho^3 I\;,
\end{align}
with
\begin{align}
I\equiv\int_0^{+\infty}x^2\left(\cos\frac{\pi}{\sqrt{1+x^{-2}}}+1\right)\approx1.104\;.
\end{align}
The total energy of the system is then
\begin{align}
\mathbb E(\rho)=8\pi^2\left(1+\frac{\rho^2}6+\frac{64}{5\Lambda^2\rho^2}\right)+32\frac{m\pi(2\Lambda\pi)^{\frac12}}{N_c}I\rho^3\;.\label{quintic}
\end{align}
As expected, the mass term works as an inertia factor, favoring
configurations of smaller radius. 

Minimization of Eq.~\eqref{quintic} cannot be done analytically. In
order to obtain some insight of the deformed quantities, we proceed by
writing $\rho=\rho_0+x$, where  
\begin{align}
\rho_0=\frac{4}{\sqrt{\Lambda}}\left(\frac3{10}\right)^{\frac14}\;,
\end{align}
and making a power series expansion of the derivative of
Eq.~\eqref{quintic} around $x=0$, truncating at first order and then  
finding the zero of the derivative (ramification point). Then, the
result can also be expanded in a power series using $m$ or $\frac{1}{N_c}$.
The deformation is  
\begin{align}
\rho=\frac{4}{\sqrt{\Lambda}}\left(\frac3{10}\right)^{\frac14}\left(1-\left(\frac65\right)^{\frac14}\frac{36I}{\sqrt\pi}\frac{m}{N_c}+{\cal O}\Big(\big(\tfrac{m}{N_c}\big)^2\Big)\right)\;.\label{size}
\end{align}
The mass of the system is equal to the energy at this value of $\rho$:
we get (after rescaling the energy) 
\begin{align}\label{baryonmass}
M_B=N_c\left(\frac{\Lambda}8+\sqrt{\frac{2}{15}}\right)+\frac{16I}{\sqrt\pi}\left(\frac65\right)^{\frac14}m\;.
\end{align}
We see that the corrections brought by the mass term can be written as
a power series in $\frac{m}{N_c}$ (rescaled by an initial power of $N_c$), so
the first correction given by the mass is subleading in $N_c$.

\subsection{Equations of motion}

We will eventually be interested in the configuration with $\hA_M=0$,
since those components arise as $\Nc^{-1}$ corrections (induced
via the Chern-Simons term by the presence of a slow motion, be it
rotational or translational), so we neglect the factor
$\exp\big(-\i\int dz \hA_z\big)$. We also still impose the condition that
the masses of the two light flavors are degenerate.
The Aharony-Kutasov action contributes to the equations of motion
with: 
\begin{align}
\frac{\delta \S_{\rm AK}}{\delta A_z} =&-\i \frac{\left(2\Lambda\pi\right)^{\frac{1}{2}}}{N_c} m\left(e^{\i\int_{-\infty}^{+\infty} dz A_z} - e^{-\i\int_{-\infty}^{+\infty} dz A_z}\right) \nonumber\\
= & -\frac{2\left(2\Lambda\pi\right)^{\frac{1}{2}}}{N_c}m \sin\left(\int_{-\infty}^{+\infty} dz A_z\right)\;.
\end{align}
Keeping only the first order of the sine power series, we obtain the
new equation of motion for the component $A_z$ (i.e. it is a set of
three equations, since $A_z = A_z^a T^a$): 
\beq
H(z)^{\frac{3}{2}}\left(\de_i\de_i A_z - \de_i\de_z A_i\right)
-\frac{2\left(2\Lambda\pi\right)^{\frac{1}{2}}}{N_c}m\int_{-\infty}^{+\infty} dz' A_z (x,z')= \text{Source terms}\;,
\eeq
while the other equations are unaltered (to be precise, the equation
of motion for $\hA_z$ would receive a correction if we go beyond the
static approximation, but that is beyond our goal to include time
derivatives only). 
Since the holonomy of the $A_z$ field is dual to the pion field, and
since the first-order term in the equations of motion arise from a
Lagrangian term quadratic in it, the modification we performed to the
equations of motion corresponds completely with the relaxation of the
massless pion regime, in favor of a finite pion mass. This is fully
consistent with the quark description: by giving finite mass to single
quarks, pions are now pseudo-Goldstone bosons, hence massive. However,
since the quark masses are degenerate, we still have a residual
symmetry, reflected in the degenerate masses of the pion
triplet.\footnote{It 
  could look like the mass of the $\eta'$ meson is also bound to be
  the same as that of the pion triplet (in this limit with only two
  flavors), since the corresponding mass term has to originate from
  this very same Aharony-Kutasov action, and the fields are equally
  normalized. However, the Chern-Simons term for the Ramond-Ramond
  form, $C_7$, induces the holographic realization of the
  Witten-Veneziano mechanism, thus removing this degeneracy. See
  Ref.~\cite{Sakai:2004cn}, and
  Refs.~\cite{Bartolini:2016dbk,Leutgeb:2019lqu} for more detailed
  discussions}

For completeness and later convenience, we provide once again the full 
set of equations of motion restoring the parity indices and the
sources:
\begin{align}\label{eomhA0}
H^{-\frac{1}{2}}\de_i\de_i\hA_0+\de_z\left(H^{\frac{3}{2}}\de_z \hA_0\right)&=-\frac{32\pi^2}{\Lambda}\delta^3(x)\delta(z)\;,\\\label{eomAi+}
H^{-\frac{1}{2}}\de_j\de_jA_i^++\de_z\left(H^{\frac{3}{2}}\de_z A_i^+\right)&=-2\pi^2 \rho^2\epsilon_{ijk}\sigma_k\de_j\delta^3(x)\delta(z)\;,\\\label{eomAz}
H^{\frac{3}{2}}\left(\de_i\de_i A_z^+ - \de_i\de_z A_i^-\right)  -a\int_{-\infty}^{+\infty} dz' A_z (x,z')&= -2\pi^2 \rho^2 \sigma_i \de_i \delta^3(x)\delta(z)\;, \\\label{eomAi-}
\frac{\de_j\de_jA_i^--\de_j\de_iA_j^-}{H^{\frac{1}{2}}}-\de_z\left(H^{\frac{3}{2}}\left(\de_iA_z^+-\de_z A_i^-\right)\right)&=2\pi^2 \rho^2 \sigma_i  \delta^3(x)\de_z\delta(z)\;,
\end{align}
where we have defined the new parameter
\beq
a \equiv \frac{2\left(2\Lambda\pi\right)^{\frac{1}{2}}}{N_c}m\;.
\eeq
We define the fields in terms of Green's functions $G(x,z,x',z')$ and
$L(x,z,x',z')$ as:
\begin{align}
\hA_0&=-\frac{32\pi^2}{\Lambda} G(x,z,0,0)\;,\nn\\
A_i^+&=-2\pi^2 \rho^2\epsilon_{ijk}\sigma_k\de_jG(x,z,0,0)\;,\nn\\
A_i^-&=-2\pi^2 \rho^2 \sigma_i  \de_{z'}G(x,z,0,z')\big|_{z'=0}\;,\nn\\
A_z^+&=-2\pi^2 \rho^2 \sigma_i \de_i L(x,z,0,0)\;,\label{AnsatzA}
\end{align}
and we take the functions $G(x,z,x',z')$ and $L(x,z,x',z')$ to obey
\begin{align}\label{eqG}
H^{-\frac{1}{2}}\de_i\de_iG(x,z,0,0)+\de_z\left(H^{\frac{3}{2}}\de_z G(x,z,0,0)\right)
&=\delta^3(x-x')\delta(z-z')\;,\\\label{eqGLz}
H^{\frac{3}{2}}\left(\de_i\de_i L(x,z,x',z')- \de_i\de_z G(x,z,x',z')\right)  \nn\\
-a\int_{-\infty}^{+\infty} dz' L(x,z,x',z')
&=  \delta^3(x-x')\delta(z-z')\;, \\ \label{eqGL}
H^{-\frac12}\de_{z'}G(x,z,x',z')+\de_z(H^{\frac32}L(x,z,x',z'))&=0\;,
\end{align}
with $G(x,z,x',z')$ and $L(x,z,x',z')$ given as in
Refs.~\cite{Baldino:2017mqq,Hashimoto:2008zw} by 
\begin{align}
G(x,z,x',z') &= -\frac{1}{4\pi}\sum_{n=1}^{\infty}\label{ansatzG} \frac{\psi_n(z)\psi_n(z')}{c_n}\frac{e^{-k_n|x-x'|}}{|x-x'|}\;,\\
L(x,z,x',z') &= -\frac{1}{4\pi}\sum_{n=0}^{\infty}	\label{ansatzL} \frac{\phi_n(z)\phi_n(z')}{d_n}\frac{e^{-k_n|x-x'|}}{|x-x'|}\;.
\end{align}

Our goal is to derive the values of $k_n$ dual to the meson masses
again, to check how the presence of finite quark masses influences the
masses of the bound states. 
To start off, we define a scalar product as in
Ref.~\cite{Baldino:2017mqq}: 
\beq\label{product()}
(f,g) \equiv \int_{-\infty}^{+\infty} dz\; H^{-\frac{1}{2}}(z)f(z)g(z)\;.
\eeq
The Ans\"atze \eqref{ansatzG}-\eqref{ansatzL} and Eq.~\eqref{eqG} give
us the usual condition 
for the profile functions $\psi_n(z)$: 
\beq
H^{\frac{1}{2}}(z)\de_z\left(H(z)^{\frac{3}{2}}\de_z \psi_n(z)\right) = -k_n^2 \psi_n(z)\;,
\eeq
so that $\psi_n(z)$ are found as solutions to an eigenvalue problem of
an Hermitian (with respect to Eq.~\eqref{product()}) operator. They must
then obey a completeness relation given by 
\begin{align}
\sum_{n=1}^{\infty} \frac{\psi_n(z)\psi_n(z')}{H(z)^{\frac12}c_n}  =
\delta(z-z')\;,  \qquad c_n = (\psi_n,\psi_n)\;, 
\end{align}
so that at this stage there are no differences compared to the
massless quarks problem. The pion profile function corresponds to 
$\phi_0$ so let us turn our attention to that: Eq.~\eqref{eqGL} is
solved by imposing, for every $n$, the conditions
\begin{align}
\frac{\de_z(H(z)^{\frac{3}{2}}\phi_n(z)) \phi_n(z')}{d_n} + \frac{\psi_n(z) \left(\de_{z'}\psi_n(z')\right)}{H(z)^{\frac{1}{2}} c_n} &= 0\;,\label{eq:GL2}\\
\frac{\de_z(H(z)^{\frac{3}{2}}\phi_0(z))\phi_0(z')}{d_0} &= 0\;.\label{eq:GLz2}
\end{align}
Eq.~\eqref{eq:GL2} is solved by choosing $\phi_n(z)=\de_z\psi_n(z)$ and
$d_n=k_n^2c_n$: note however that $d_0$ is not determined by this
condition, since there is no normalizable mode $\psi_0$. Finally,
Eq.~\eqref{eq:GLz2} imposes the shape of the pion wave function
$\phi_0(z)=H^{-\frac{3}{2}}(z)$.	 

Now we substitute into Eq.~\eqref{eomAz} obtaining
\beq\nonumber
H(z)^{\frac{3}{2}} \sum_{n=0}^{\infty} \frac{\phi_n(z)\phi_n(z')}{d_n}\delta^3(x-x')
+\frac{a}{4\pi} \sum_{n=0}^{\infty} \frac{\int_{-\infty}^{+\infty}dz\phi_n(z)\phi_n(z')}{d_n} \frac{e^{-k_n |x-x'|}}{|x-x'|}\nn\\
\mathop- H(z)^{\frac{3}{2}}\frac{1}{4\pi} \frac{\phi_0(z)\phi_0(z')}{d_0}k_0^2\frac{e^{-k_0|x-x'|}}{|x-x'|}=\delta^3(x-x')\delta(z-z')\;.\label{eqnazansatz}
\eeq
Here all contributions to the sum in the second term vanish due to the
relation $\phi_n(z) = \de_z \psi_n(z)$ and the normalizability of the
eigenfunctions $\psi_n(z)$: 
\beq
\int_{-\infty}^{+\infty}dz\phi_n(z) = \psi_n(+\infty) - \psi_n(-\infty)\;, \qquad n\geq 1\;,
\eeq
so that only the $n=0$ contribution survives. In the first term we can
make use of a completeness relation for the functions $\phi_n(z)$: We
define a new inner product 
\beq
\langle f,g\rangle \equiv \int_{-\infty}^{+\infty} dz\; H(z)^{\frac{3}{2}} f(z) g(z) \; ,
\eeq
which satisfies $\langle\phi_n,\phi_n\rangle=d_n=k_n^2 c_n$ for all
$n$ but $n=0$. Extending the notation to $\phi_0$ and performing the
integration we identify $d_0 = \pi$ and obtain the completeness
relation 
\begin{equation}
\sum_{n=0}^{\infty} \frac{H(z)^{\frac32}\phi_n(z)\phi_n(z')}{d_n} = \delta(z-z')\; ,\qquad d_n = \langle\phi_n,\phi_n\rangle\;.
\end{equation}
Exploiting this new relation, we can check that the first term of
Eq.~\eqref{eqnazansatz} cancels out with the deltas on the right hand
side.
Hence all we are left with is the following equation: 
\begin{equation}
\frac{a}{4\pi} \phi_0(z') \frac{e^{-k_0 |x-x'|}}{|x-x'|} - \frac{H(z)^{\frac{3}{2}}}{4\pi}\frac{\phi_0(z)\phi_0(z')}{d_0}k_0^2\frac{e^{-|x-x'|}}{|x-x'|}=0\;,
\end{equation}
that we can further simplify by recalling that $\phi_0(z) =
H^{-\frac{3}{2}}(z)$. After doing so, it is possible to identify the
pion mass by recalling that the $k_n^2$ are dual to the masses of the
mesons, so we obtain 
\beq\label{pionmass}
k_0^2 = \pi a = \frac{\sqrt{(2\pi)^3\Lambda}}{N_c} m\;.
\eeq

An alternative, perhaps more intuitive way to obtain the same result
is to obtain the effective Lagrangian for the mesons by expanding the
gauge fields: doing so for the Aharony-Kutasov action, and remembering
that the holonomy of the $\mA_z$ field is related to the pseudoscalars
as  
\begin{equation}
\mU \equiv \mathcal{P}\exp\left(-\i\int dz\mA_z \right) = \exp\left(\frac{\i}{f_{\pi}}(\pi^a\tau^a + S\mathds{1})\right)\;,
\end{equation}
we can easily derive the Gell-Man--Oakes--Renner relation for this
model: 
\beq
4mc = f_{\pi}^2 m_{\pi}^2\;,
\eeq
from which the pion mass squared can be read off.
Making use of Eq.~\eqref{LAMBDA} and the holographic formula for the
pion decay constant 
\beq
f_{\pi}^2 = 4\frac{\kappa}{\pi}\;,
\eeq
it is easy to verify that $m_{\pi}^2$ indeed coincides with $k_0^2$ of
Eq.~\eqref{pionmass}.

Indeed, moving to higher orders in derivatives, further terms are
induced in the action by the presence of the quark mass: in chiral 
perturbation theory a term can be written that has the form
\beq
\frac{1}{f_{\pi}^2}\tr\left(MU D_{\mu}UD^\mu U\right)\;,
\eeq 
which is the leading correction to the $\tr\left(MU\right)$ term, and 
generates mass terms for (axial) vector mesons. However, as
argued in Ref.~\cite{Hashimoto:2009hj}, this term is holographically
realized as a correction of relative order $\Lambda^{-2}$, hence we
neglect this contribution in the usual $1/\Lambda$ expansion.


\section{Nucleon-Nucleon potential}
\label{quattro}

We now turn to computing the interaction potential between two
nucleons: this is done along the lines of the previous work
(i.e.~Ref.~\cite{Baldino:2017mqq}). As a first step we build a
two-instanton configuration by positioning two instantons at a
distance far larger than their size and giving them arbitrary
orientations: 
\beq\label{twoansatz}
\hA= \hA^p + \hA^q\;, \qquad A = BA^pB^{\dag} + CA^q C^\dag\;,
\eeq
with $\big(X_{(p)},Z_{(p)}\big) = (0,0,0,0)$ and $\big(X_{(q)},Z_{(q)}\big) = (r^1,r^2,r^3,0)$.
In general, this field configuration is not a solution to the
equations of motion, but it is an approximation at leading order in
$\Lambda$ in the limit $R\gg2\rho$ with $R$ being the distance between
the instanton centers. This is a viable approximation scheme since
the holographic model extrapolates from the large $\lambda$ regime,
and the size of the single instantons is of order
$\rho^{-\frac{1}{2}}$ while $R\simeq \lambda^{0}$. Space is then
divided into three regions: two balls of radius $\rho$ centered at
$\vec{X}=\vec{X}_{(p)}$ and $\vec{X}=\vec{X}_{(q)}$, where the
corresponding field is strong and the other is in its linear
approximation, and the rest of space where both fields can be
approximated by their linear form. 

To compute the (static) energy, we employ the definition
$\S = -\int dt \mathcal{E}$ so that
\beq  \label{energydef}
\mathcal{E} = \kappa^{-1} E =- \kappa^{-1} \int d^3x\,dz\;\mathcal{L}\;,
\eeq
is the rescaled energy, while $E$ is the energy in physical units
($\mathcal{L}$ is the Lagrangian density in physical units). Part of
the integral in Eq.~\eqref{energydef} will account for the self energies
of the solitons (the masses of the single baryons), so to compute the
interaction potential, we have to subtract off these self energy
terms: the Ansatz \eqref{twoansatz} allows us to easily do so by
keeping just the cross-terms that involve both fields $A^{p,q}$. 
The full rescaled energy is obtained by the on-shell action as:
\begin{align} 
\mathcal{E} &= \int d^3x\,dz\bigg(\frac{1}{2H^{\frac{1}{2}}}\tr\left(F_{ij}^2\right) + H^{\frac{3}{2}}\tr\left(F_{iz}^2\right)
+ \frac{1}{2H^{\frac{1}{2}}} \big(\de_i \hA_0\big)^2 + \frac{H^{\frac{3}{2}}}{2}\big(\de_z\hA_0\big)^2 \bigg)\nn\\
&\phantom{=\ }
\mathop+a\int d^3x\,dz\;\tr\left(\int_{-\infty}^{z} dz'\, A_z(x,z)A_z(x,z')\right)\;, \label{energy}
\end{align}
where we have made use of the equations of motion to trade the Chern
Simons term for a change in sign in the $\hA_0$ terms of the
Yang-Mills action.  

Now it is sufficient to insert Eq.~(\ref{AnsatzA}) into the above
expression and exploit Eqs.~(\ref{eqG})-(\ref{eqGL})
together with the Ans\"atze (\ref{ansatzG}) and (\ref{ansatzL}) to be
able to perform integrations with Dirac deltas. The only difference
that emerges with respect to Ref.~\cite{Baldino:2017mqq}, is the
presence of the last term in Eq.~(\ref{energy}), and that now
$k_0\neq 0$. 
We introduce the symmetric tensor:
\begin{equation}
P_{ij}(r,k) = \delta_{ij}\left((rk)^2 +rk+1\right) -\frac{r_ir_j}{r^2}\left((rk)^2+ 3rk +3\right),
\end{equation}
as well as the rotation matrix
\beq
M_{ij} (G) = \frac{1}{2}\tr\left(\sigma_i G \sigma_j G^\dag\right) ,
\label{eq:M_rotation}
\eeq
which gives the spatial rotation corresponding to an SU(2) rotation
implemented via the matrix $G\in$ SU(2). 
The full potential at the end of a somewhat lengthy computation (but analogous to
that of section 3.1 of Ref.~\cite{Baldino:2017mqq}) is given in
physical units by:
\beq
V(r,B^\dag C) =\frac{4\pi N_c}{\Lambda}\left( \sum_{n=1}^{\infty}\left(\frac{1}{c_{2n-1}}\frac{e^{-k_{2n-1}r}}{r}+\frac{6}{5}\frac{1}{c_{2n-1}}M_{ij}(B^\dag C)P_{ij}(r_i,k_{2n-1})\frac{e^{-k_{2n-1}r}}{r^3}\right.\right. \nonumber \\
\left.\mathop-\frac{6}{5}\frac{1}{d_{2n}}\frac{e^{-k_{2n}r}}{r^3}M_{ij}(B^\dag C)P_{ij}(r_i,k_{2n})   \right) -\frac{6}{5\pi} \frac{e^{-k_0 r}}{r^3}M_{ij}(B^\dag C) P_{ij}(r_i,k_0)\Bigg) \; .
\eeq
In this formula, $r_i$ is the relative position of the single soliton
cores, while $r^2=r_ir_i$. We see immediately that the difference
with respect to the potential obtained in Ref.~\cite{Baldino:2017mqq}
is entirely contained in the last term, corresponding to the
contribution of the $n=0$ mode, that is, the pion: here we have simply
another short-range interaction (actually absorbable in the previous
term by extending the sum to $n=0$ and remembering that $d_0=\pi$),
mediated by a particle of mass $k_0$, as can be argued by the
exponential decay. In Ref.~\cite{Baldino:2017mqq} we had a long-range
interaction, as appropriate for a massless mediating boson.

Here we assumed fixed a position in the $z$ coordinate and fixed
the instanton sizes: It is possible to consider these quantities as
massive moduli as long as we consider small oscillations around their
equilibrium value, so that they will actively enter in the expression
of the potential to give the more general formula 
\begin{align}
&V(r,B^\dag C, \rho_i, Z_i)\nn\\ &=\frac{4\pi N_c}{\Lambda} \sum_{n=1}^{\infty}\frac{\psi_{n}(Z_1)\psi_{n}(Z_2)}{c_{n}}\frac{e^{-k_{2n-1}r}}{r}+\frac{\pi}{16}N_c\Lambda \rho_1\rho_2\frac{\phi_0(Z_1)\phi_0(Z_2)}{\pi}\frac{e^{-k_0 r}}{r^3}M_{ij}(B^\dag C)P_{ij}(r_i,k_0)\nn\\
&\quad \ +\frac{\pi}{16}N_c\Lambda \rho_1\rho_2\left(\sum_{n=1}^{\infty}\left(\frac{\psi_n(Z_1)\psi_n(Z_2)}{c_n}+\frac{\phi_n(Z_1)\phi_n(Z_2)}{d_n}\right)\frac{e^{-k_n r}}{r^3}P_{ij}(r_i,k_n)\right)M_{ij}(B^\dag C)\;.
\label{potential}
\end{align}
As we can see now, every sum runs over all values of $n$, not
alternating between even and odd values: It is easy to verify that
this generalization is due to the presence of $Z_i\neq 0$, and that in
the case $Z_i=0$ the appropriate terms vanish because of the parity
properties of the functions $\psi_n(z)$, and $\phi_n(z)$.

\section{Numerical evaluations}
\label{cinquebis}

\subsection{Fit of the free parameters}

We now proceed to the numerical evaluation of the various physical
quantities that we have found. In our model, we have three free
parameters\footnote{In principle, the number
	of colors, $N_c$, is also a free parameter, but we will make the
	obvious choice $N_c=3$. }, $M_{KK},\Lambda$ and $m$, of which
$M_{KK}$ is the only dimensional quantity.
We calibrate the parameters to the following values: 
\begin{itemize}
	\item For the mass parameter $M_{KK}$, we choose the value $949$
	MeV. This is chosen to fit the mass of the $\rho$ meson, that is the
	second-lightest massive meson in our model. Even if the usual
	practice is to fit the mass scale using the lightest massive
	particle in the model, we choose to use the $\rho$ meson for the fit,
	following Ref.~\cite{Baldino:2017mqq}. This way the introduction of
	the pion mass will appear numerically in the model as a
	perturbation of the massless model.
	\item For the coupling $\Lambda$, we choose the value $1.569$. This is
	done to fix the pion decay constant in the massless model. 
	\item The new parameter $m$ is interpreted as the quark mass. We have
	chosen the value $3.066\times10^{-3}$ for this parameter. This
	yields a quark mass of $2.910$ MeV, which is between the masses of
	the $u$ and $d$ quarks. 
\end{itemize}
With the above parameter values, we get the following values for the
observables:  
\begin{itemize}
	\item Pion mass: Evaluating Eq.~\eqref{pionmass}, we get a pion
	mass of $134.8$ MeV. This is a very good result, as the
	experimental pion masses are $135.0$ MeV for the neutral pion
	($\pi_0$) and $139.6$ MeV for the charged pions
	($\pi_{\pm}$). This result confirms that the model works well at
	low energies.
	\item Baryon mass: Evaluating Eq.~\eqref{baryonmass}, we get a baryon
	mass of $1.628$ GeV for the light baryons (neutrons and
	protons). The baryons in our model are significantly heavier than
	the physical baryons, whose masses sit around $938$ MeV for the proton
	and neutron. This is to be expected, as the model is expected to
	break down at energy scales larger around $949$ MeV. 
\end{itemize}

\subsection{Composite nuclei}

We will now turn to the minimization of the 2-body potential acting on
$B$ nuclei as
\beq
V_{\rm tot}^B = \sum_{\substack{i,j=1\\i\neq j}}^B V(r_{ij},B_i^\dag B_j)\;,
\eeq
where $r_{ij}\equiv |x_i-x_j|$ and $B_i$ is the SU(2) rotation matrix
of the $i$-th nucleon (instanton).
The numerical methods we will use are a simple random walk algorithm
(only moving forward when the potential energy is lowered) and a
Metropolis algorithm allowing for statistical random movement (with
some probability of moving ``uphill'') as well as a simple gradient
flow method.

\begin{figure}[!ht]
\centering
\includegraphics[width=\linewidth]{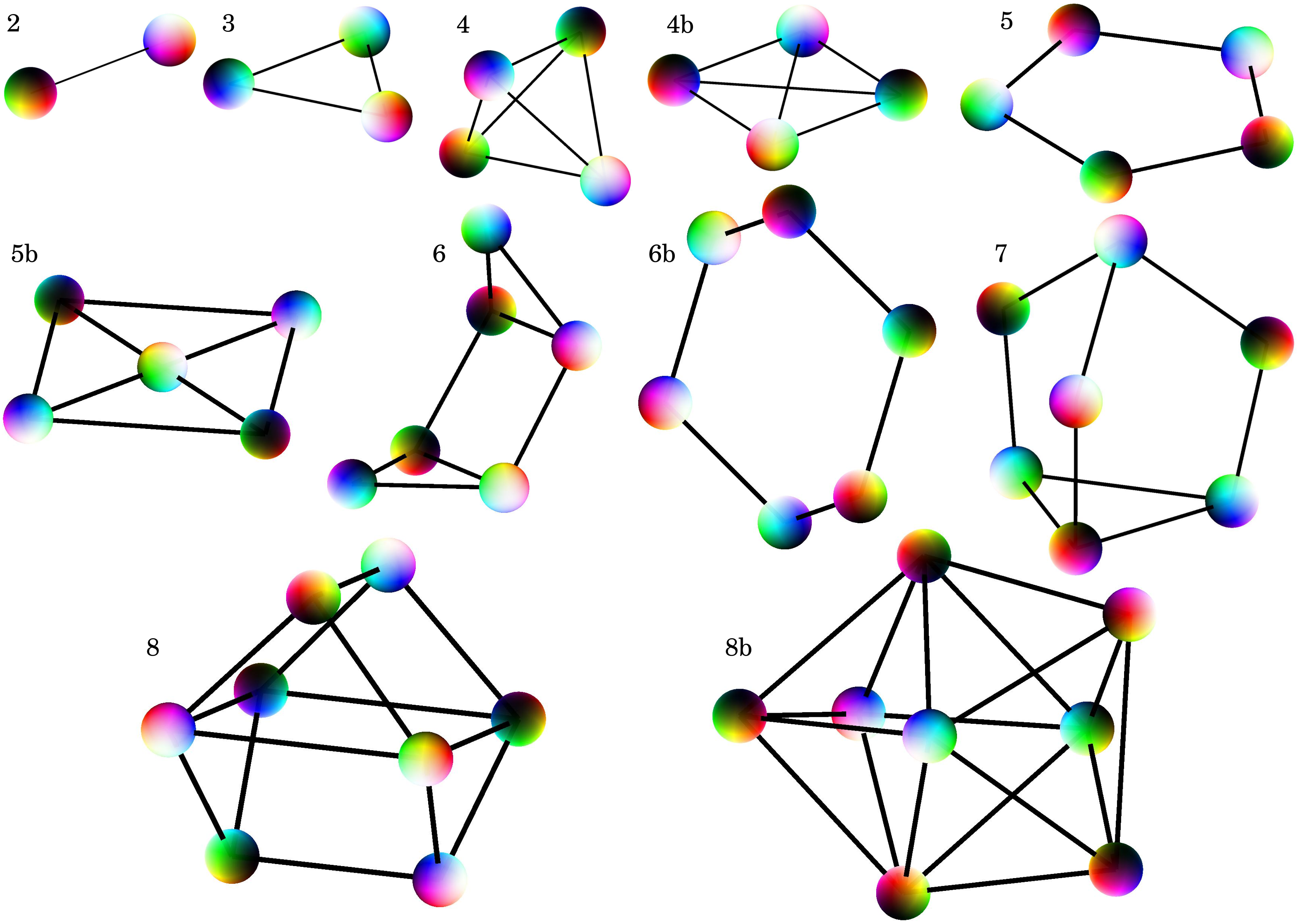}
\caption{Geometric configurations for stable and metastable nuclei up
  to $B=8$.
  The stable (energetically favorable) configurations are labeled with
  the baryon number $B$ and the metastable states have an added
  suffix in form of a Latin letter; the further in the alphabet the
  higher the energy. 
  Bonds are shown for short enough distances between the
  nuclei as black solid lines.
  The color scheme is described in the text.
}
\label{minima}
\end{figure}
In Fig.~\ref{minima} we present the numerical results for classical
composite nuclei, stable and meta-stable, up to baryon number $B=8$.
The stable configurations (ground states) are labeled with their
corresponding baryon number $B$, while metastable states have a Latin
letter added to the label. With increasing letter in the alphabet, the
higher is the energy.
This bond between two nuclei symbolizes the 2-body potential, which is
used to find these configurations and the bond is displayed only for
distances shorter than $1.5R_0$ with $R_0$ being the
optimal distance of two nucleons in deuterium ($B=2$) for
$m_\pi=0.142$.
The color scheme utilized is a map from the unit 2-sphere to the
Runge's color sphere, which is defined by white at the north pole,
black at the south pole and the hue of the colors around the equator,
going through red, yellow, green, cyan, blue, and magenta as one goes
around the equator.
The color scheme illustrates the orientation of the point-like
instantons in SU(2) space, with the Runge's color sphere being the
standard orientation corresponding to the unit matrix
$B=\mathds{1}_2$.
All points on the color sphere for a nonstandard orientation of the
instanton are then rotated by the rotation matrix $M_{ij}(B)$ of
Eq.~\eqref{eq:M_rotation}.

\begin{figure}[!ht]
  \centering
  \includegraphics[width=0.6\linewidth]{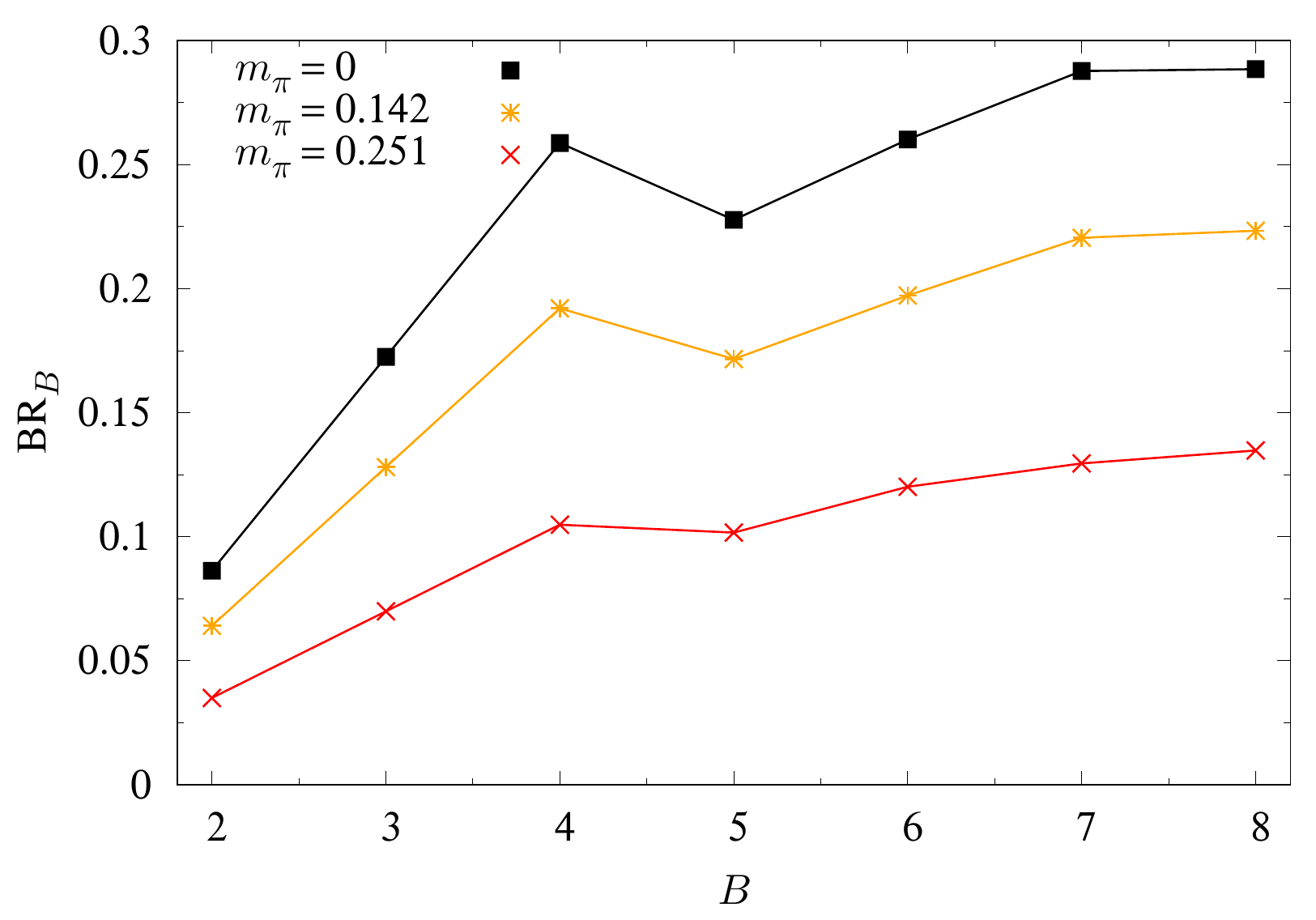}
  \caption{Binding ratios for the stable nuclei (ground states) up to
    $B=8$, for various values of $m$. In our calibration, the physical
    pion mass is $m_\pi=0.142$ (dimensionless units).
  }
  \label{fig:BR}
\end{figure}
We will now calculate the binding ratios of the $B$ nuclear bound
states, defined by
\beq
{\rm BR}_B \equiv \frac{BE_1 - E_B}{BE_1}\;.
\eeq
The result is shown in Fig.~\ref{fig:BR}.
Qualitatively, the bound states show the same spatial/geome\-trical
distributions in the case of a finite pion mass as compared to the
massless case, see Fig.~\ref{minima} and Fig.~5 of
Ref.~\cite{Baldino:2017mqq}.
The optimal distance between two nucleons in the 2-body potential
grows with an increasing pion mass and this effectively enlarges all  
the sizes of the nuclear bound states, but does not affect the
orientations of the instantons in SU(2) space.
In particular, the optimal distance of the 2-body potential for
$m_\pi=0$ in dimensionless units is $R_0=2.06$ and for $m_\pi=0.142$ 
it is $R_0=2.23$. 
The binding energies and thus the binding ratios, similarly decrease
with an increasing pion mass.

On the other hand, the presence of the pion mass removes the
long-range force of the potential, so even though the minimum is
pushed a bit away in the 2-body potential, the attraction is also
shortened somewhat.
This has the consequence that some states are different in the massive
case, with respect to the massless case.
In particular, a metastable state in the $B=5$ sector made of a
tetrahedron with a satellite exists in the massless case
\cite{Baldino:2017mqq}, but has disappeared in the massive case.
Similarly, the ground state of the $B=6$ state has changed by moving
the leftmost and the topmost nucleon in Fig.~\ref{minima} closer to
the hexagon in the massive case, than in the massless case where it
was more compact.
\begin{table*}[!ht]
  \begin{center}
    \begin{tabular}{c|p{5cm}|p{5cm}|c|c}
      $B$ & shape & details & $V_{\rm min}$ & BR\\\hline
      2 & line & bond length = 2.23 & $-0.2197$ & $0.0640$\\\hline
      3 & equilateral triangle & bond length = 2.23 & $-0.6590$ & $0.1280$\\\hline
      4 & tetrahedron & bond length = 2.23 & $-1.3179$ & $0.1920$\\
      & square & side bond length = 2.07& $-1.2499$ & $0.1821$\\\hline
      5 & pentagon & bond length = 2.25 & $-1.4721$ & $0.1716$\\
      & cross & side bond length = 3.20 & $-1.4432$ & $0.1682$\\\hline
      6 & two triangles forming a bent hexagon, triangles bent the
      same way & triangle bond lengths = 2.13, 2.28; square bond
      length = 3.34 & $-2.0304$ & $0.1972$\\
      & two triangles forming a bent hexagon, triangles bent the
      opposite way & bond length = 2.17 & $-2.0023$ & $0.1945$\\\hline
      7 & tetrahedron + triangle & bond lengths = 2.12, 2.46, 2.80 & $-2.6486$ & $0.2205$\\\hline
      8 & two triangular prisms sharing the same base, but rotated by
      90 degrees & bond lengths = 2.47, 2.70 & $-3.0660$ & $0.2234$\\
      & pyramid + triangle & bond lengths = 2.96, 3.06, 3.15, 3.21, 3.24 & $-2.5576$ & $0.1863$\\
    \end{tabular}
    \caption{Details of stable and metastable multi-instanton
      solutions for $B=2,\ldots8$. $V_{\rm min}$ is the potential in
      dimensionless units and BR is the binding ratio. }
    \label{tab:details}
  \end{center}
\end{table*}
The details of the solutions in the massive case with $m_\pi=0.142$
are shown in Tab.~\ref{tab:details}.


\section{The deuteron state}
\label{sei}

The quantum description of the deuteron is obtained by quantization of
the collective modes starting from the $B=2$ classical system that we
examined in section \ref{quattro}. The procedure has already been
described in detail in Ref.~\cite{Baldino:2017mqq}: Here we will review the
analysis and obtain the binding energy of the deuteron in the system
at hand. 

A generic field configuration in the two-instanton sector is written
by considering only the degrees of freedom that do not change the
potential computed in section \ref{quattro}, and fixing all other
degrees of freedom to have the potential as attractive as
possible. The unfixed degrees of freedom form the zeromode manifold,
and can be interpreted in the following way: 
\begin{itemize}
\item Overall position of the center of mass of the system, that we
  denote as $x=(x^1,x^2,x^3)$. Due to the gravitational field in the
  $z$-direction, we do not allow motion along the $z$-axis. Quantizing
  those degrees of freedom gives the system an overall momentum. 
\item Isospin of the system: This is represented by an SU(2) matrix
  called $U$. 
\item Spin of the system: This is represented by an SU(2) matrix
  called $E$ and the corresponding SO(3) matrix, $M_{ij}(E)$ is given
  in Eq.~\eqref{eq:M_rotation}. 
\item Parity: We can always switch the two components of the
  system. As this is a discrete symmetry, it will not induce a
  momentum. We will indicate parity with a binary variable $P=0,1$. 
\end{itemize}
The relative position and phase are minimized by assuming maximal
attraction: This is attained at the separation $R=(R_0,0,0)$ (with both
nuclei centered at the origin of the holographic coordinate) with phase
opposition, $B^\dag C=\i\sigma_3$.
The numerical value of $R_0$ is the value at which the potential
\eqref{potential} is minimized. 

A configuration $\mathcal A_I$ in the two-instanton sector belongs to
the zeromode manifold, if it can be written as 
\begin{align}
\mathcal A_I(x,z)&=UE^\dagger A_I\left(x-(-)^PM(E)\frac r2,z\right)(UE^\dagger)^\dag\nn\\&\mathop+U\i\sigma_3 E^\dagger A_I\left(x+(-)^PM(E)\frac r2,z\right)(U\i\sigma_3 E^\dagger)^\dagger\;.
\label{zeromode}
\end{align}
The kinetic energy is computed by starting from the kinetic energy of
two baryons, computed in Ref.~\cite{Sakai:2004cn}, and then comparing
configuration \eqref{zeromode} with configuration \eqref{twoansatz} to
understand how the collective coordinates are related to the
coordinates describing the single baryons. After performing the
transformation and freezing the coordinates that are not zeromodes,
we obtain the kinetic energy for the zeromode manifold. Since the
derivation is the same as in Ref.~\cite{Baldino:2017mqq}, we will just
cite the result here. In terms of the angular left-invariant
velocities $\Omega_i=-\i\tr(U^\dagger\dot U\sigma_i)$ and 
$\omega_i=-\i\tr(E^\dagger\dot E\sigma_i)$, we have the kinetic energy 
\beq
T=\frac12 M_B\bigg(\rho^2\omega_1^2+\left(\rho^2+\frac{R_0^2}2\right)\omega_2^2+\frac{R_0^2}2\omega_3^2+\rho^2\left(\Omega_1^2+\Omega_2^2+(\Omega_3-\omega_3)^2\right)\bigg)\;.
\eeq
The mass $M$ and radius $\rho$ have been computed in subsection
\ref{deformation}, and the parameter $R_0$ is obtained from the
potential \eqref{potential} in the attractive channel, by finding the
minimum.
The Hamiltonian for the deuteron is computed straightforwardly. First,
the momenta are defined as
\begin{align}
L_1&=M_B\rho^2\omega_1\;,&\quad
L_2&=M_B\left(\rho^2+\frac{R_0^2}2\right)\omega_2\;,&\quad
L_3&=M_B\left(\frac{R_0^2}2+\rho^2\right)\omega_3-M\rho^2\Omega_3\;,\nn\\
K_1&=M_B\rho^2\Omega_1\;,&\quad
K_2&=M_B\rho^2\Omega_2\;,&\quad
K_3&=M_B\rho^2(\Omega_3-\omega_3)\;.
\end{align}
In terms of those momenta, the Hamiltonian reads
\bea
H=\frac1{2M_B}\left(X_{ij}L_iL_j+Y_{ij}K_iK_j+2Z_{ij}L_iK_j\right)+V_{\rm min}+2M_B\;,
\eea
where we have included rest mass and defined the inertia tensors
$X,Y,Z$ as 
\begin{align}
X&\equiv
\left(\begin{array}{ccc}
\frac1{\rho^2}&0&0\\
0&\frac{2}{ 2\rho^2 +R_0^2}&0\\
0&0&\frac2{R^2_0}\\
\end{array}\right) ,\\ \qquad 
Y&\equiv
\left(\begin{array}{ccc}
\frac1{\rho^2}&0&0\\
0&\frac{1}{\rho^2}&0\\
0&0&\frac2{R_0^2} + \frac1{\rho^2}\\
\end{array}\right) ,\\
Z&\equiv
\left(\begin{array}{ccc}
0&0&0\\
0&0&0\\
0&0&\frac2{R_0^2}\\\end{array}\right) .
\end{align}
Quantum states are expressed as
\begin{align}
\ket\psi=\ket{k,k_3,i_3,l,l_3,j_3},
\end{align}
where $k$ and $l$ are eigenvalues for $K^2$ and $L^2$, of values
$\frac k2(\frac k2+1)$ and $\frac l2(\frac l2+1)$, respectively, $k_3$
and $l_3$ are eigenvalues for $K_3$ and $L_3$, and $i_3$ and $j_3$ are
eigenvalues for the right-invariant momenta $I_3$ and $J_3$, that
commute with the (left-invariant) momenta $K_3$ and $L_3$ and have the
properties $I^2=K^2$ and $L^2=J^2$. The right-invariant momenta can be
obtained from the left-invariant momenta as $J_i=-M(E)_{ij}L_j$ and
$I_i=-M(U)_{ij}K_j$ (where $E$ and $U$ are promoted to position
operators). 

Due to the fact that the configuration space has discrete symmetries
(as an example, multiplying $U$ and $E$ by $\i\sigma_1$ on the right
leaves the configuration invariant) we have to impose
Finkelstein-Rubenstein constraints. The analysis in
Ref.~\cite{Baldino:2017mqq} is still valid: The only states that are
compatible with the quantization of the single baryons as fermions and
that include the constraints are 
\begin{align}
\ket D&=\ket{0,0,0,1,0,j_3},\qquad \ket {I_0}=\ket{1,0,i_3,0,0,0},\nn\\
\ket{I_1}&=\frac1{\sqrt2}\left(\ket{1,1,i_3,0,0,0}+\ket{1,-1,i_3,0,0,0}\right).
\end{align}
Of those states, $\ket D$ is the only one having the quantum numbers
of the deuteron (isospin $0$ and spin $1$). The energies of the
states are computed by acting on them with the Hamiltonian: the final
result is
\begin{align}
\nonumber H\ket{D} &=  \left(\frac{1}{2\rho^{2}M_B}\left(1+\frac{1}{1+\frac{R_{0}^{2}}{2\rho^{2}}}\right)+2M_B+V_{\rm min}\right)\ket{D}=E_D\ket{D}\;,\\\nonumber 
H\ket{I_{0}} &=  \left(\frac{1}{\rho^{2}M_B}+2M_B+V_{\rm min}\right)\ket{I_0}=E_{I_0}\ket{I_0}\;,\\ 
H\ket{I_{1}} &=  \left(\frac{1}{\rho^{2} M_B}\left(1+\frac{\rho^{2}}{R_{0}^{2}}\right)+2M_B+V_{\rm min}\right)\ket{I_{1}}=E_{I_1}\ket{I_1}\;.
\label{rote}
\end{align}
It is evident that $E_D$ is always the smallest energy of the three
ones above, so the deuteron state is effectively the ground state of
the system. Although the result is identical to the result in
Ref.~\cite{Baldino:2017mqq} in form, we emphasize that in this setting
the values of $M_B,\rho,V_{\rm min}$ and $R_0$ are different, so the final
numerical result will be different.

With $E_D$ obtained from the first line of (\ref{rote}), the most relevant quantity we can compute is the binding energy $E_D-2M_B$.
Various parameters are needed to obtain this value. The
new parameters $R_0$ and $V_{\rm min}$ are not free, but they are fixed by
looking for the minimum of the classical potential \eqref{potential}
in the attractive channel ($B^\dagger C=\sigma_3,
r=(R_0,0,0),\rho_i=\rho,Z_i=0$). In the massive model, we have found
$R_0=2.23$ and $V_{\rm min}=-0.220$ by including $40$ massive mesons in the
potential computation. Furthermore, the instanton size can be computed
from Eq.~\eqref{size}, obtaining a value of $\rho=2.307$. We can
combine these parameters to obtain the binding energy. The classical
binding energy in MeV is given by $-V_{\rm min}=208.5$ MeV, and the
rotational corrections have a small influence on this value: the
quantum binding energy is given by $E_D-2M_B=208.4$ MeV. This is to
be compared against a binding energy of $2.224$ MeV found in
experiments. As in our previous work, the binding energy is two orders
of magnitude larger than the expected binding energy. The pion mass
improves the prediction, as the binding energy of the massless model
is $275.5$ MeV. A study of the $\frac{1}{N_c}$ and $\frac{1}{\Lambda}$ corrections is
required to understand if the prediction for the binding energy can be 
improved, as the parameters $N_c$ and $\Lambda$ are not that large.

\section{An infinite lattice of baryons}
\label{sette}

The attractive channel of the potential between two nucleons is
obtained by employing a relative rotation between their SU(2)
orientations: The iso-rotation must act by an angle $\pi$ with an axis
orthogonal to the relative position vector. Using the axis-angle
notation, the matrix $M_{ij}(B^\dag C)$ that describes the relative
orientation is given by 
\begin{equation}\label{axisangle}
M_{ij}(\hat{\bu},\alpha) = \delta_{ij}\cos\alpha + \left(1-\cos\alpha\right)\hat{u}_i\hat{u}_j + \epsilon_{ijk} \hat{u}_k \sin\alpha\;.
\end{equation}
\begin{figure}[!htp]
  \begin{center}
    \includegraphics[width=0.5\linewidth]{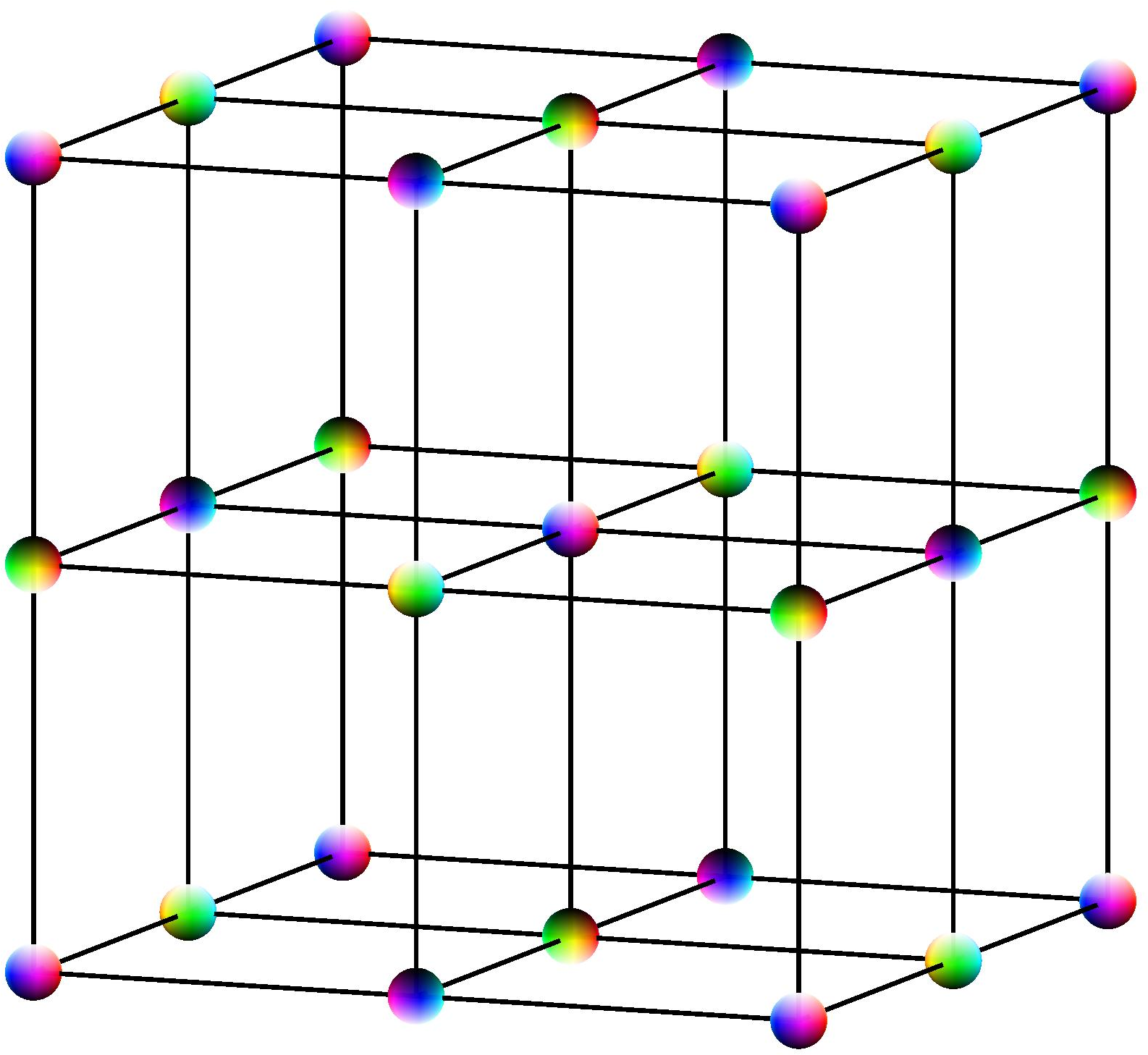}
    \caption{The fundamental cell of the infinite lattice. Careful
      inspection of the figure reveals that every pair of nucleons,
      connected with a solid black line, face each other with a
      matching color, representing the attractive channel described in
      the text.
    }
    \label{fig:lattice}
  \end{center}
\end{figure}
It is possible to arrange nucleons in a cubic lattice in such a way
that the interaction between all nearest neighbors is in the attractive
channel, see fig.~\ref{fig:lattice}.
The Ansatz for the lattice we are using is equal to that of
Ref.~\cite{Klebanov:1985qi}.
The relation between the instanton lattice studied here and the
Skyrmion lattice of Ref.~\cite{Klebanov:1985qi} is given by the
holonomy of the instanton gauge field \cite{Atiyah:1989dq}.
We make the following choices: The relative iso-orientation
$B^\dag C$ between every soliton and its nearest neighbors is given
by 
\begin{itemize}
\item $\pm\i\sigma^2$ for $\br=(\pm R,0,0)$, corresponding to $\alpha=\pi,\hat{\bu}=\hat{x}_2$,
\item $\pm\i\sigma^3$ for $\br=(0,\pm R,0)$, corresponding to $\alpha=\pi,\hat{\bu}=\hat{x}_3$,
\item $\pm\i\sigma^1$ for $\br=(0,0,\pm R)$, corresponding to $\alpha=\pi,\hat{\bu}=\hat{x}_1$.
\end{itemize}
This choice for the nearest neighbors fixes completely the lattice in a
self-consistent way (since $\sigma^1\sigma^2=\i\sigma^3$), so it is now
possible to compute the energy density associated to it. 

Before moving to the computation, let us make some useful
considerations: The relative orientation enters the potential formula
via the combination $M_{ij}P_{ij}(\br,k)$, which reads 
\begin{align}
  M_{ij}(\hat{\bu},\alpha)P_{ij}(\br,k) &= \bigg(1+\cos\alpha-\left(1-\cos\alpha\right)\frac{(\hat\bu\cdot\br)^2}{r^2}\bigg)(kr)^2\nn\\&\phantom{=\ }-(1-\cos\alpha)\left(3\frac{(\hat\bu\cdot\br)^2}{r^2}-1\right)\left(rk+1\right)\;,  \qquad
  r\equiv \sqrt{\br\cdot\br}\;,\label{MP}
\end{align}
so that we have two classes of iso-orientations: $\alpha=0$ and $\alpha=\pi$:
\begin{align}
M_{ij}(\hat{\bu},0)P_{ij}(\br,k) & =2(rk)^2\nn\\
M_{ij}(\hat{\bu},\pi)P_{ij}(\br,k) &=2\frac{(\hat\bu\cdot\br)^2}{r^2}\left[-3(rk+1)-(rk)^2\right]+2(rk+1)\;.
\end{align}
To compute the full energy density it is sufficient to evaluate the
potential between a single chosen nucleon and all the others: Then
translational symmetry ensures that the full energy is given by
multiplying this contribution by the number of nucleons (which is
infinite) with a factor of one half to avoid double counting. The
nucleon number can then be traded for the number of cells, as a
function of the lattice volume $V_L$, so the full energy will be
computed as 
\beq
E = \frac{1}{2}\sum_p \sum_{p'\neq p} V(\br_{pp'},B_p^\dag C_{p'}) = \frac{1}{2}\frac{V_L}{R^3}\sum_{p'\neq p} V(\br_{pp'},B_p^\dag C_{p'})\;, \qquad
\br_{pp'} \equiv \br_{p'} - \br_p\;,
\eeq
with $p$ indicating a lattice site.

To make the calculation more straightforward, we choose our reference
frame in such a way that $\br_p = (0,0,0)$ and $B_p = \mathds{1}$,
so we are left with the following formula for the energy density:  
\begin{equation}
\frac{E}{V_L} = \frac{1}{2R^3}\sum_{p'}V(\br_{p'},C_{p'})
=\frac{1}{2R^3}\sum_{p'}V(\br_{p'},\alpha_{p'},\hat{\bu}_{p'})\;,
\end{equation}
where $p'$ runs over all lattice sites with the exception of $(0,0,0)$.
The lattice sites are uniquely parametrized as
\beq
\br=R(n_1,n_2,n_3)\;,\qquad n_i \in \mathbb{Z}\;.
\eeq
The position $\br$ not only enters the expression of $M_{ij}P_{ij}$,
but it also determines whether $\alpha=0$ or $\alpha=\pi$: We must
therefore be careful and classify iso-orientations with respect to
$(n_1,n_2,n_3)$. There are four possible scenarios, depending on how
many even or odd instances of $n_i$ are present in the position
vector: 
\begin{itemize}
	\item all $n_i$ are EVEN $\Rightarrow C_{p'} = \mathds{1}$,
	\item all $n_i$ are ODD $\Rightarrow C_{p'} = \mathds{1}$,
	\item one $n_i$ is EVEN $\Rightarrow C_{p'} = \pm\i\sigma^j$, 
	\item one $n_i$ is ODD $\Rightarrow C_{p'} = \pm\i\sigma^j$,
\end{itemize}
where in the last two lines $\sigma^j$ indicates a rotation
around the axis determined by our rule for the rotation of nearest
neighbors (that is, $j=2$ for $i=1$, $j=3$ for $i=2$ and $j=1$ for
$i=3$).
For example, a nucleon sitting at $R(2b_1-1,2b_2,2b_3)$ with
$b_i\in\mathbb{Z}$ will have an orientation of $\pm\i\sigma^2$, i.e.~the
same as a nucleon sitting at $R(2b_1,2b_2-1,2b_3-1)$. The last two
scenarios fall into the class of $\alpha=\pi$, and the axis of
rotation enters Eq.~\eqref{MP} only via $(\hat\bu\cdot\br)$, which
selects the component of $\br$ along the axis $\hat{\bu}$. But since
$\hat{\bu}$ is a unit vector corresponding to one of the coordinate
axes, this simply selects the $j$-th component $R n_j$.

Now we will compute the contributions to the energy density:
Let us start with the first two possibilities in the classification,
corresponding to all even or all odd $n_i$, for which we obtain
\begin{align}
  &\frac{E_{1,2}}{V_L} = \frac{4\pi N_c}{2R^3\Lambda}\!\sum_{b_1,b_2,b_3}\!\!\left[\sum_{n=1}^{\infty}\left(\frac{1}{c_{2n-1}}\frac{e^{-k_{2n-1}r}}{r}+\frac{12k_{2n-1}^2}{5c_{2n-1}}\frac{e^{-k_{2n-1}r}}{r}-\frac{12k_{2n}^2}{5d_{2n}}\frac{e^{-k_{2n}r}}{r}  \right)  -\frac{12k_0^2}{5\pi}\frac{e^{-k_0r}}{r}\right]\!,
  \label{eq:E12}
\end{align}
with $r^2 = 4R^2(b_1^2 +b_2^2+b_3^2)$ for the even case ($E_1$) and
$r^2 = R^2[(2b_1-1)^2 +(2b_2-1)^2+(2b_3-1)^2]$ for the odd
case ($E_2$).
We now turn to the more difficult scenarios: We start by considering
$\br=(2b_1-1,2b_2,2b_3)R$ with $b_i\in\mathbb{Z}$. In this case the
energy density becomes
\beq
\frac{E_3^{(1)}}{V_L}  = \frac{E_{3,4}^{p}(4b_2^2)}{V_L}\;,
\eeq
where we have defined
\begin{align}
 \frac{E_{3,4}^{p}(b)}{V_L} \equiv  & \frac{4\pi N_c}{2R^3\Lambda}\sum_{b_1,b_2,b_3}\left[ \sum_{n=1}^{\infty}
\left[\frac{1}{c_{2n-1}}\frac{e^{-k_{2n-1}r}}{r}\right.\right.\nn\\
&\nonumber\left.\left.+\frac{6}{5}\frac{2}{c_{2n-1}}\left(\frac{b}{r^2} \left(-3(rk_{2n-1}+1)-(rk_{2n-1})^2\right)+\left(rk_{2n-1}+1\right)\right)\frac{e^{-k_{2n-1}r}}{r^3}
\right.\right. \nonumber \\
&\nonumber\left.\left.-\frac{6}{5}\frac{2}{d_{2n}}\left(\frac{b}{r^2} \left(-3(rk_{2n}+1)-(rk_{2n})^2\right)+\left(rk_{2n}+1\right)\right)\frac{e^{-k_{2n}r}}{r^3}   \right]\right. \\
&-\frac{12}{5\pi} \frac{e^{-k_0 r}}{r^3}\left(1+rk_0 -\frac{bR^2}{r^2}\left(3+3k_0r+k_0^2r^2\right)\right)\Bigg]\;,
\label{eq:E34}
\end{align}
with $r^2 = R^2[(2b_1-1)^2 +4b_2^2+4b_3^2]$.
We can see that the other terms labeled by $E^{(2,3)}_3$ will have the
same form except for the substitution of every instance of $b_2$ with
respectively $b_3,b_1$: Since the sum runs over all three parameters,
then each of these terms will give contribution equal to the result 
\beq
\frac{E_3}{V_L} = 3\frac{E_3^{(1)}}{V_L}\;.
\eeq
We are thus only left with the terms of the case in which one
coordinate is an even multiple of the lattice spacing, while the
others are odd. As before we begin by analyzing the situation in which
this coordinate is $r_1$, so that $\br=R(2b_1,2b_2-1,2b_3-1)$: The
result is just as the previous one, with the exception that the
projection on the $j$-th component of $\br$ due to the rotation axis
will now pick up an odd value instead of an even one, so that 
\beq
\frac{E_4^{(1)}}{V_L} = \frac{E_{3,4}^{p}((2b_2-1)^2)}{V_L}\;.
\eeq
and $r$ is now given by
$r^2 = R^2[4b_1^2 +(2b_2-1)^2+(2b_3-1)^2]$.
Again, as in the previous case, the sum of all three possible
combinations with only one even coordinate will just result in a
factor of three, so we get
\beq
\frac{E_4}{V_L} = 3\frac{E_4^{(1)}}{V_L}\;.
\eeq
The total energy density is then given by the sum of the contributions
from all four possibilities in the classification, giving
\beq\label{totinteraction}
\frac{E_{\rm tot}}{V_L} = \frac{1}{V_L}\left(E_1+E_2+E_3+E_4\right)\;.
\eeq

\begin{figure}[!ht]
  \centering
  \includegraphics[width=0.6\linewidth]{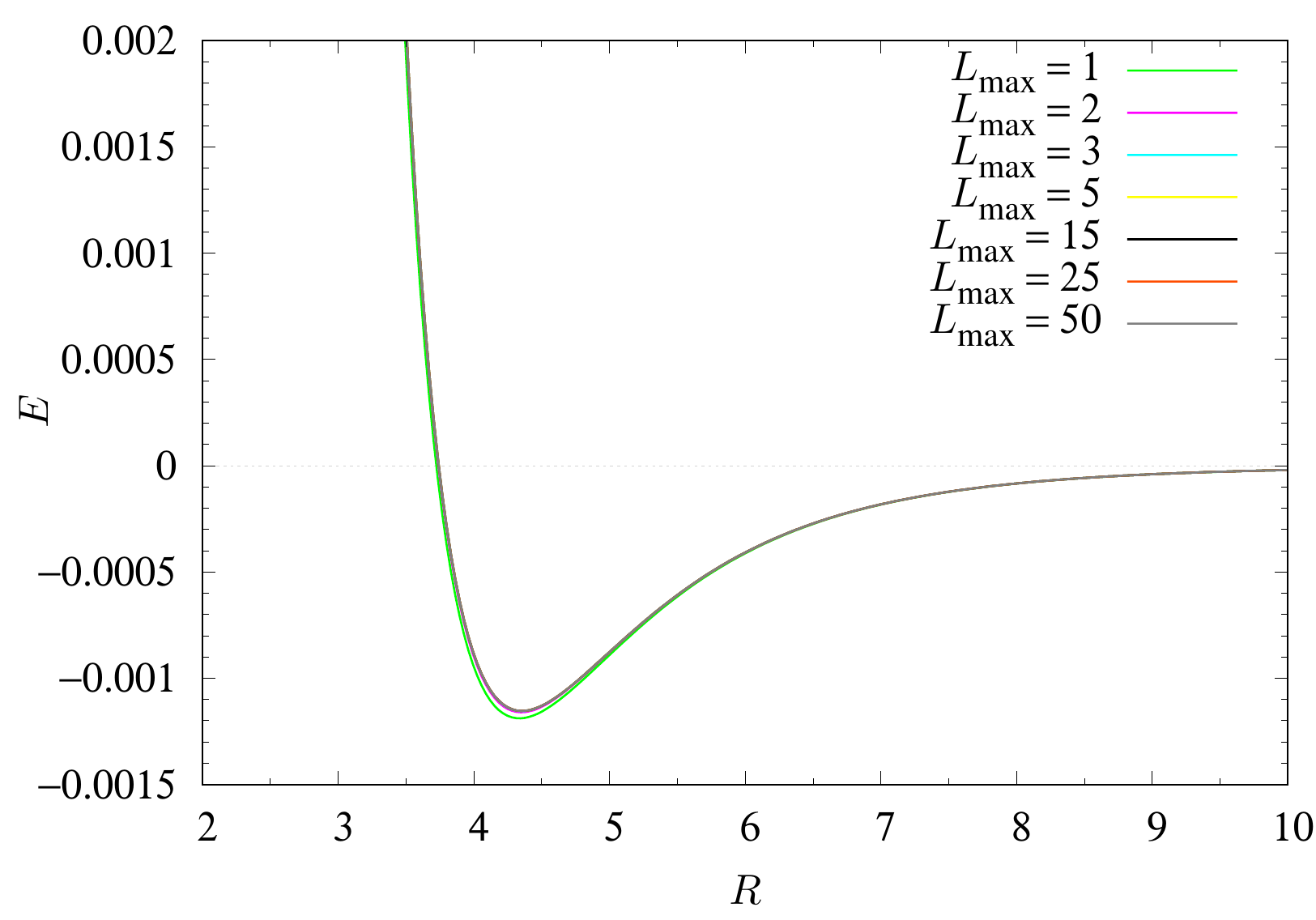}
  \caption{The shape of the potential energy density around its
    minimum. $L_{\rm max}$ is the cutoff of lattice sites:
    $|b_j|\leq L_{\rm max}$. }
  \label{potentialminimum}
\end{figure}
The total lattice potential can then be computed numerically and
plotted as a function of the lattice spacing $R$: the sum over lattice
sites converges with good precision ($\sim 10^{-5}$) for
$b_j \in [-L_{\rm max},L_{\rm max}]$ with $L_{\rm max}=15$, $\forall$
$j=1,2,3$.
Notice that the cutoff in terms of lattice sides is $2L_{\rm max}$ in
each direction (i.e.~both in the $x+$ direction and in the $x-$ direction)
and the factor of two is due to $b_j$ being an integer parametrizing
even or odd $n_j$'s. 
The resulting potential density has a shallow minimum at $R=4.86$ (see
Fig.~\ref{potentialminimum}). 

It is not straightforwardly clear that the quadruple sum in
Eqs.~\eqref{eq:E12} and \eqref{eq:E34} are convergent in the limit of
$L_{\rm max}\to\infty$, i.e.~in the limit of summing over all lattice
sites $b_j$ and all massive vectors $n$ (being the index of $k_n$).
The convergence of the quadruple sum would prove that the density in the
given form is finite.
We give a mathematical proof of the convergence of each term in the
sums separately in appendix \ref{app:convergence}.

\subsection{Hadronic phase transition}

We will now use the baryon lattice to study the presence of a hadronic
phase transition: To do so we need to compute the free energy of the
configuration and look for a critical density at which it becomes
negative. 

We can do this without relying on holography, simply by calculating
the Legendre transform of the energy density. First of all, we recover
the total energy density from the interaction potential density by adding
the mass density terms: the cubic lattice cell has unit net baryon
number, so the mass density as a function of the lattice spacing $R$ is
simply 
\beq
\mathcal{M}(R) = M_B R^{-3}\;,
\eeq
with $M_B$ given by Eq.~\eqref{baryonmass}.

The chemical potential is defined as the derivative of the energy
density with respect to the baryon number density (which we can trade
for a derivative with respect to $R$, since the baryon number density
is $d_B=R^{-3}$):
\beq
\mu = -\frac{R^{4}}{3}\frac{\de}{\de R}\left(EV_L^{-1} + M_B R^{-3}\right),
\eeq
with $EV_L^{-1}$ given by Eq.~\eqref{totinteraction}. 

We want to compare the free energy density of the infinite
lattice of baryons constructed above, with that of the vacuum:
the embedding of the flavor branes is the same for both phases, so we
neglect its contribution to both phases, effectively setting to zero
the free energy of the vacuum. All we need to compute is then the
change in free energy due to the presence of baryons: the lattice
configuration will become favored over the vacuum (that is, a purely
mesonic phase) for negative values of this free energy.
The free energy density $F$ is then obtained from the energy density
as: 
\begin{align}
  F &= EV_L^{-1} + \mathcal{M}(R) - \mu R^{-3} \nn\\
  &= EV_L^{-1} + \frac{R}{3}\frac{\de}{\de R}EV_L^{-1}\;.
\end{align}

\begin{figure}[!ht]
  \centering
  \includegraphics[width=0.6\linewidth]{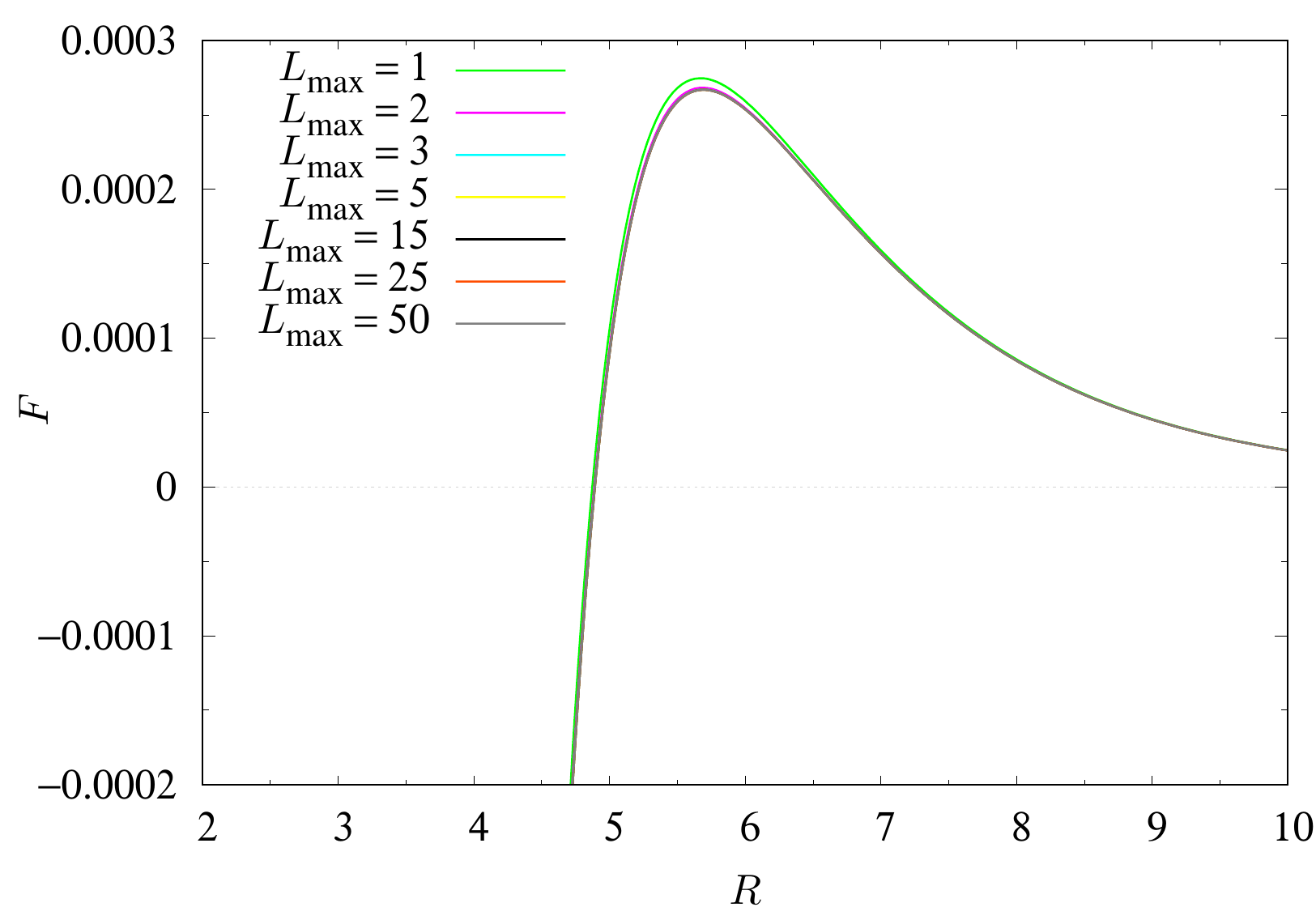}
  \caption{The free energy density as a function of the lattice
    spacing $R$. As the density increases (i.e.~$R$ decreases), the
    free energy becomes negative, signaling a phase
    transition. $L_{\rm max}$ is the cutoff of lattice sites: 
    $|b_j|\leq L_{\rm max}$. }
  \label{freeenergy}
\end{figure}

In Fig.~\ref{freeenergy}, we plot the free energy density as a
function of $R$, showing that it becomes negative at a finite density,
signaling the presence of a first-order hadronic phase
transition.
Note that with this analysis we cannot conclude that the
favored of all phases is indeed given by this lattice of baryons, but
only that it is favored over the vacuum. Different configurations of
baryonic matter can potentially give rise to even lower free energies,
and become favored over this particular lattice.

\section{Conclusion}
\label{conclusion}

In this paper, we have included the quark mass term using the
Aharony-Kutasov action in our solitonic approach to holographic
nuclear physics.
We work in the limit $\Lambda\to\infty$, where the size of the
instantons is much smaller than the typical separation distance
between nuclei in a nuclear bound state.
This allows us to calculate multi-instanton solutions by gluing
together two overlapping instantons in the linear regime; in
particular, we compute the 2-body potential which now has acquired a
mass term for the pions.
This induction of the pion mass from the quark mass term is in line
with the GMOR relation.
Using numerical methods we find nuclear bound states -- stable and
metastable -- with baryon numbers $B=2$ through $B=8$.

We find that the main difference by having massive pions in the 2-body
potential is that the nuclear bound states grow slightly in size and
correspondingly reduce their binding energy.
A similar effect applies also to the deuteron bound state and to the
other light nuclei.

Using the 2-body potential, we consider the case of an infinite
crystal of nucleons in a cubic lattice with each nucleon oriented in
SU(2) space so as to put it in the attractive channel with respect to
nearest neighbors.
We calculate the potential energy density, prove in appendix 
\ref{app:convergence} that it is finite, and finally use it to
calculate the free energy.
The free energy becomes negative at a critical lattice spacing
$R_{\rm crit}\sim4.9$, which signals a hadronic phase transition of
first order. The presence of the hadronic phase transition was also
studied in the same model in
Refs.~\cite{Li:2015uea,Preis:2016gvf,BitaghsirFadafan:2018uzs,Kovensky:2019bih}:
The difference with our approach lies both in the description of the
solitons and in the setup of the flavor branes. In the cited works the
branes' position at infinity are taken to be close enough to allow the
use of the deconfined geometry up to arbitrarily low temperatures,
thus being far from the antipodal branes' regime: the baryonic matter
is described with various levels of accuracy, starting from exactly
pointlike instantons up to an infinite lattice of finite size
instantons interacting with the nearest neighbors via the ADHM
construction. We work instead with antipodal branes at zero
temperature, using the confined geometry, and we employ the flat space
instanton approximation, then take into account every single
interaction (up to a cutoff in lattice size) with every soliton seeing
each other's core as pointlike. The presence of the first order phase
transition remains a 
feature of the model in both regimes, and is becoming of second order if
interactions between instantons are ignored \cite{Li:2015uea}.

The work carried out in this paper has been done in the large
$\Lambda$ and large $\Nc$ limit, which is of course none other than
approximations to real world physics, as neither parameters take
(that) large phenomenological values (i.e.~$\Lambda_{\rm SS}=1.569$
and $\Nc=3$).
It would be interesting to calculate, for instance, the leading
$\Lambda^{-1}$ correction to the 2-body potential and see if this
could improve some phenomenological properties of our results, for
example the large binding energies.
Of course, this would turn the problem into a nonlinear one and make
it severely more difficult than the one we have solved here.

The large-$N_c$ approximation may be a poor approximation for the
physics at finite density, such as the baryon crystal studied in the
last part of this paper, and subleading $1/N_c$ corrections may alter
our results and even the presence or absence of phase transitions, see 
e.g.~Refs.~\cite{Torrieri:2010gz,Lottini:2011zp}.
This can be attributed to the fact that the number of nearest
neighbors in a cubic lattice is 6 and $N_c=3$ in Nature, but in the
large-$N_c$ limit, the number of colors is much larger than the number
of nearest neighbors -- this has consequences for, e.g.~the van Der
Waals forces, etc.

\section*{Acknowledgments}

S.~Baldino wishes to dedicate this work to the memory of Federico
Tonielli, to remember an extraordinary person and an excellent
scientist who left us prematurely. 
The work of S.~Baldino is supported by FCT - Funda\c\~ao para a
Ci\^encia e a Tecnologia, through the PhD fellowship
SFRH/BD/130088/2017.
The work of L.~Bartolini is supported by the
``Fondazione Angelo della Riccia''.
The work of S.~Bolognesi is supported by the INFN special project
grant ``GAST (Gauge and String Theories)''.  
S.~B.~Gudnason thanks the Outstanding Talent Program of Henan
University for partial support. 
The work of S.~B.~Gudnason is supported by the National Natural
Science Foundation of China (Grants No.~11675223 and No.~12071111).

\appendix
\section{Proof of convergence}
\label{app:convergence}

We now give a proof that the energy density for the infinite lattice
of baryons  (\ref{totinteraction}) is a convergent sum, by giving an
upper bound for each term.
In this appendix, $m$ denotes a parameter comparable to the
pion mass and should not be confused with the quark mass in the main
text.
\begin{lemma}
  The following quadruple sum obeys the inequality
  \beq
  \sum_{\substack{n=0\\b_1,b_2,b_3=-\infty\\(b_1,b_2,b_3)\neq0}}^{\infty}
  \frac{e^{-k_n R\sqrt{b_1^2+b_2^2+b_3^2} }}{c_n R^p(b_1^2+b_2^2+b_3^2)^{\frac{p}{2}}}
  \leq\frac{\calA_p(m,R)\calG{\left(\frac{\eta R}{2}\right)}+\calB_p(m,R)\calG{\left(\frac{3\eta R}{2}\right)}}{\min(\{c_n\})}\;,
  \eeq
  provided $k_n$ obeys $k_n\geq m+\eta n$ with $m>0$, $\eta>0$ positive
  constants, and $p\in\mathbb{Z}_{>0}$.
  \label{lemma:1}
\end{lemma}
\emph{Proof}:
Using $
\sqrt{A^2+B^2+C^2} \geq \frac{|A|+|B|+|C|}{2}$, 
 we have
for the triple sum
\begin{align}
\sum_{\substack{b_1,b_2,b_3=-\infty\\(b_1,b_2,b_3)\neq (0,0,0)}}^{\infty}
e^{-k\sqrt{b_1^2+b_2^2+b_3^2}} \leq
\sum_{\substack{b_1,b_2,b_3=-\infty\\(b_1,b_2,b_3)\neq (0,0,0)}}^{\infty}
e^{-\frac{k}{2}|b_1|}
e^{-\frac{k}{2}|b_2|}
e^{-\frac{k}{2}|b_3|}
=\frac{3e^{\frac{k}{4}}+e^{-\frac{3k}{4}}}{4\sinh^3{\left(\frac{k}{4}\right)}}\;.
\end{align}
At this point we include the fourth sum over $n$:
\begingroup
\allowdisplaybreaks
\begin{align}
&\sum_{\substack{n=0\\b_1,b_2,b_3=-\infty\\(b_1,b_2,b_3)\neq (0,0,0)}}^{\infty}
e^{-k_nR\sqrt{b_1^2+b_2^2+b_3^2}} \leq\sum_{\substack{n=0\\b_1,b_2,b_3=-\infty\\(b_1,b_2,b_3)\neq (0,0,0)}}^{\infty}
e^{-(m+\eta n)R\sqrt{b_1^2+b_2^2+b_3^2}}\non
&\ \ \leq\sum_{\substack{n=0\\b_1,b_2,b_3=-\infty\\(b_1,b_2,b_3)\neq (0,0,0)}}^{\infty}
e^{-\frac12(m+\eta n)R(|b_1|+|b_2|+|b_3|)}=\sum_{n=0}^\infty\frac{3e^{\frac14(m+\eta n)R}+e^{-\frac34(m+\eta n)R}}{4\sinh^3{\left(\frac14(m+\eta n)R\right)}}\non
&\ \ \leq\frac{3e^{\frac{mR}{4}}\calG{\left(\frac{\eta R}{2}\right)}+e^{-\frac{3mR}{4}}\calG{\left(\frac{3\eta R}{2}\right)}}{4\sinh^3{\left(\frac{mR}{4}\right)}}\;,
\label{eq:fourthsum}
\end{align}
\endgroup
where we have used that
\beq
\sinh{\left(\frac{m R}{4}+\frac{\eta n R}{4}\right)}
\geq\sinh{\left(\frac{m R}{4}\right)}e^{\frac{\eta n R}{4}}\;, \quad
m>0\;, \quad R>0\;, \quad n\in\mathbb{Z}_{\geq 0}\;,
\eeq
and we have defined
\beq
\calG(x) \equiv \frac{1}{1 - e^{-x}}\;, \qquad x>0\;.
\eeq
Next we need  the Yukawa-like sum
\begin{align}
\sum_{\substack{b_1,b_2,b_3=-\infty\\(b_1,b_2,b_3)\neq (0,0,0)}}^{\infty}
\frac{e^{-k R\sqrt{b_1^2+b_2^2+b_3^2}}}{R^p(b_1^2+b_2^2+b_3^2)^{\frac{p}{2}}}
&=\int_k^\infty \d{m_1}\cdots \int_{m_{p-1}}^\infty \d{m_p}
\sum_{\substack{b_1,b_2,b_3=-\infty\\(b_1,b_2,b_3)\neq (0,0,0)}}^{\infty}
e^{-m_pR\sqrt{b_1^2+b_2^2+b_3^2}}\non
&\leq\int_k^\infty \d{m_1}\cdots \int_{m_{p-1}}^\infty \d{m_p}
\left[\coth^3{\left(\frac{m_p R}{4}\right)}-1\right],
\end{align}
for $p\in\mathbb{Z}_{>0}$ and $k>0$.
Combining this result with the fourth sum of Eq.~\eqref{eq:fourthsum},
we obtain the following result
\begin{align}
&\sum_{\substack{n=0\\b_1,b_2,b_3=-\infty\\(b_1,b_2,b_3)\neq (0,0,0)}}^{\infty}
\frac{e^{ -k_n R\sqrt{b_1^2+b_2^2+b_3^2} }}{R^p(b_1^2+b_2^2+b_3^2)^{\frac{p}{2}}} \leq\sum_{\substack{n=0\\b_1,b_2,b_3=-\infty\\(b_1,b_2,b_3)\neq (0,0,0)}}^{\infty}
\frac{e^{-(m+\eta n) R\sqrt{b_1^2+b_2^2+b_3^2}}}{R^p(b_1^2+b_2^2+b_3^2)^{\frac{p}{2}}}\non
&\ \ =\int_m^\infty \d{m_1}\cdots \int_{m_{p-1}}^\infty \d{m_p}
\sum_{\substack{n=0\\b_1,b_2,b_3=-\infty\\(b_1,b_2,b_3)\neq (0,0,0)}}^{\infty}
e^{-(m_p+\eta n) R\sqrt{b_1^2+b_2^2+b_3^2}}\non
&\ \ \leq\int_m^\infty \d{m_1}\cdots \int_{m_{p-1}}^\infty \d{m_p}
\frac{3e^{\frac{mR}{4}}\calG{\left(\frac{\eta R}{2}\right)}+e^{-\frac{3mR}{4}}\calG{\left(\frac{3\eta R}{2}\right)}}{4\sinh^3{\left(\frac{mR}{4}\right)}}\non
&\ \ =\calA_p(m,R)\calG{\left(\frac{\eta R}{2}\right)}+\calB_p(m,R)\calG{\left(\frac{3\eta R}{2}\right)}\;,
\end{align}
with the functions
\begingroup
\allowdisplaybreaks
\bea
&   \calA_1(m,R)  = \frac{6\big(1-2e^{\frac{m R}{2}}\big)}{R\big(e^{\frac{m R}{2}}-1\big)^2},\quad  \calB_1(m,R)  = \frac{2}{R}\left[m R + \frac{3-2e^{\frac{m R}{2}}}{\big(e^{\frac{m R}{2}}-1\big)^2} - 2\log\big(e^{\frac{m R}{2}}-1\big)\right],\nn \\
 &  \calA_2(m,R) = \frac{6}{R^2}\left[m R + \frac{2}{e^{\frac{m R}{2}}-1} - 2\log\big(e^{\frac{m R}{2}}-1\big)\right],\nn \\
&   \calB_2(m,R) = \frac{2}{R^2}\left[-3m R + \frac{2}{e^{\frac{m R}{2}}-1} + 6\log\big(e^{\frac{m R}{2}}-1\big) + 4\Li_2\big(e^{-\frac{m R}{2}}\big)\right],\nn \\
&  \calA_3(m,R) = \frac{24}{R^3}\left[\frac{m R}{2} - \log\big(e^{\frac{m R}{2}}-1\big) + \Li_2\big(e^{-\frac{m R}{2}}\big)\right],\nn \\
&   \calB_3(m,R)  = \frac{8}{R^3}\bigg[\frac{m R}{2} - \log\big(e^{\frac{m R}{2}}-1\big) - 3\Li_2\big(e^{-\frac{m R}{2}}\big)
  + 2\Li_3\big(e^{-\frac{m R}{2}}\big)\bigg]\;,
\eea
\endgroup
where $\Li_2$ is the dilogarithm or Spence function and $\Li_3$ is the
trilogarithm.
Finally, we can write
\begin{align}
\sum_{\substack{n=0\\b_1,b_2,b_3=-\infty\\(b_1,b_2,b_3)\neq (0,0,0)}}^{\infty}
\frac{e^{-k_n R\sqrt{b_1^2+b_2^2+b_3^2}}}{c_n R^p(b_1^2+b_2^2+b_3^2)^{\frac{p}{2}}}
&\leq\sum_{\substack{n=0\\b_1,b_2,b_3=-\infty\\(b_1,b_2,b_3)\neq (0,0,0)}}^{\infty}
\frac{e^{-k_n R\sqrt{b_1^2+b_2^2+b_3^2}}}{\min(\{c_n\})R^p(b_1^2+b_2^2+b_3^2)^{\frac{p}{2}}}\non
&=\frac{\calA_p(m,R)\calG{\left(\frac{\eta R}{2}\right)}+\calB_p(m,R)\calG{\left(\frac{3\eta R}{2}\right)}}{\min(\{c_n\})}\;.
\end{align}
\hfill$\square$
\begin{lemma}
  The following quadruple sum obeys the inequality
  \begin{align}
  \sum_{\substack{n=0\\b_1,b_2,b_3=-\infty\\(b_1,b_2,b_3)\neq0}}^{\infty}
  \frac{k_ne^{-k_n R\sqrt{b_1^2+b_2^2+b_3^2}}}{c_n R^{p-1}(b_1^2+b_2^2+b_3^2)^{\frac{p-1}{2}}} 
  &\leq m_+\frac{\calA_{p-1}(m_-,R)\calG{\left(\frac{\eta_- R}{2}\right)}+\calB_{p-1}(m_-,R)\calG{\left(\frac{3\eta_- R}{2}\right)}}{\min(\{c_n\})}\non
  &\phantom{\leq\ }
  -\frac{\eta_+R}{2}\frac{\calA_{p}(m_-,R)\calG'{\left(\frac{\eta_- R}{2}\right)}+\calB_{p}(m_-,R)\calG'{\left(\frac{3\eta_- R}{2}\right)}}{\min(\{c_n\})}\;,
  \end{align}
  provided $k_n$ obeys $k_n\geq m_-+\eta_-n$ and $k_n\leq m_++\eta_+n$
  with $m_\pm>0$, $\eta_\pm>0$ positive 
  constants, and $p\in\mathbb{Z}_{>1}$.
  \label{lemma:2}
\end{lemma}
\emph{Proof}: Following the lines of the proof of Lemma \ref{lemma:1},
we have 
\begin{align}
 &  \sum_{\substack{n=0\\b_1,b_2,b_3=-\infty\\(b_1,b_2,b_3)\neq0}}^{\infty}
  \frac{k_ne^{-k_n R\sqrt{b_1^2+b_2^2+b_3^2}}}{c_n R^{p-1}(b_1^2+b_2^2+b_3^2)^{\frac{p-1}{2}}}
  \leq\sum_{\substack{n=0\\b_1,b_2,b_3=-\infty\\(b_1,b_2,b_3)\neq0}}^{\infty}
  \frac{(m_++\eta_+n)e^{ -(m_-+\eta_-n) R\sqrt{b_1^2+b_2^2+b_3^2} }}{c_n R^{p-1}(b_1^2+b_2^2+b_3^2)^{\frac{p-1}{2}}}\non
  &\ \ \leq\sum_{\substack{n=0\\b_1,b_2,b_3=-\infty\\(b_1,b_2,b_3)\neq0}}^{\infty}
  m_+\frac{e^{ -(m_-+\eta_-n) R\sqrt{b_1^2+b_2^2+b_3^2} }}{c_n R^{p-1}(b_1^2+b_2^2+b_3^2)^{\frac{p-1}{2}}}
  -\sum_{\substack{n=0\\b_1,b_2,b_3=-\infty\\(b_1,b_2,b_3)\neq0}}^{\infty}
  \eta_+\frac{\d}{\d\eta_-}{\left(\frac{e^{\left(-(m_-+\eta_-n) R\sqrt{b_1^2+b_2^2+b_3^2}\right)}}{c_n R^{p}(b_1^2+b_2^2+b_3^2)^{\frac{p}{2}}}\right)}\non
  &\ \ \leq 
m_+\frac{\calA_{p-1}(m_-,R)\calG{\left(\frac{\eta_- R}{2}\right)}+\calB_{p-1}(m_-,R)\calG{\left(\frac{3\eta_- R}{2}\right)}}{\min(\{c_n\})} \non
  & 
 \ \ \ \  \ \ -\frac{\eta_+R}{2}\frac{\calA_{p}(m_-,R)\calG'{\left(\frac{\eta_- R}{2}\right)}+\calB_{p}(m_-,R)\calG'{\left(\frac{3\eta_- R}{2}\right)}}{\min(\{c_n\})}\;,
\end{align}
where $k_n\geq m_-+\eta_-n$ and $k_n\leq m_++\eta_+n$ and the
derivative with respect to the argument of $\calG$ is  
\beq
\calG'(x) = - \frac{e^{-x}}{(1 - e^{-x})^2}\;.
\label{eq:calGprime}
\eeq
\hfill$\square$
\begin{lemma}
  The following quadruple sum obeys the inequality
  \begin{align}
    \sum_{\substack{n=0\\b_1,b_2,b_3=-\infty\\(b_1,b_2,b_3)\neq0}}^{\infty}
  &\frac{k_n^2e^{-k_n R\sqrt{b_1^2+b_2^2+b_3^2}}}{c_n R^{p-2}(b_1^2+b_2^2+b_3^2)^{\frac{p-2}{2}}}   \leq m_+^2\frac{\calA_{p-2}(m_-,R)\calG{\left(\frac{\eta_- R}{2}\right)}+\calB_{p-2}(m_-,R)\calG{\left(\frac{3\eta_- R}{2}\right)}}{\min(\{c_n\})}\non
  &  -m_+\eta_+R\frac{\calA_{p-1}(m_-,R)\calG'{\left(\frac{\eta_- R}{2}\right)}+\calB_{p-1}(m_-,R)\calG'{\left(\frac{3\eta_- R}{2}\right)}}{\min(\{c_n\})} \non
  &
  +\frac{\eta_+^2R^2}{4}\frac{\calA_{p}(m_-,R)\calG''{\left(\frac{\eta_- R}{2}\right)}+\calB_{p}(m_-,R)\calG''{\left(\frac{3\eta_- R}{2}\right)}}{\min(\{c_n\})}\;,
  \end{align}
    provided $k_n$ obeys $k_n\geq m_-+\eta_-n$ and $k_n\leq m_++\eta_+n$
  with $m_\pm>0$, $\eta_\pm>0$ positive 
  constants, and $p\in\mathbb{Z}_{>2}$.
  \label{lemma:3}
\end{lemma}
\emph{Proof}: Following the lines of the proof of Lemmata \ref{lemma:1}
and \ref{lemma:2}, we have 
\begingroup
\allowdisplaybreaks
  \begin{align}
&\sum_{\substack{n=0\\b_1,b_2,b_3=-\infty\\(b_1,b_2,b_3)\neq0}}^{\infty}
  \frac{k_n^2e^{-k_n R\sqrt{b_1^2+b_2^2+b_3^2}}}{c_n R^{p-2}(b_1^2+b_2^2+b_3^2)^{\frac{p-2}{2}}}
   \leq\sum_{\substack{n=0\\b_1,b_2,b_3=-\infty\\(b_1,b_2,b_3)\neq0}}^{\infty}
  \frac{(m_++\eta_+n)^2e^{-(m_-+\eta_-n) R\sqrt{b_1^2+b_2^2+b_3^2}}}{c_n R^{p-2}(b_1^2+b_2^2+b_3^2)^{\frac{p-2}{2}}}\non
  &\ \ \leq\sum_{\substack{n=0\\b_1,b_2,b_3=-\infty\\(b_1,b_2,b_3)\neq0}}^{\infty}
  m_+^2\frac{e^{-(m_-+\eta_-n) R\sqrt{b_1^2+b_2^2+b_3^2}}}{c_n R^{p-2}(b_1^2+b_2^2+b_3^2)^{\frac{p-2}{2}}} \non
  &\phantom{=\ }
  -\sum_{\substack{n=0\\b_1,b_2,b_3=-\infty\\(b_1,b_2,b_3)\neq0}}^{\infty}
  2m_+\eta_+\frac{\d}{\d\eta_-}{\left(\frac{e^{-(m_-+\eta_-n) R\sqrt{b_1^2+b_2^2+b_3^2}}}{c_n R^{p-1}(b_1^2+b_2^2+b_3^2)^{\frac{p-1}{2}}}\right)}\non
  &\phantom{=\ }
  \ \ \ \ +\sum_{\substack{n=0\\b_1,b_2,b_3=-\infty\\(b_1,b_2,b_3)\neq0}}^{\infty}
  \eta_+^2\frac{\d^2}{\d\eta_-^2}{\left(\frac{e^{-(m_-+\eta_-n) R\sqrt{b_1^2+b_2^2+b_3^2}}}{c_n R^{p}(b_1^2+b_2^2+b_3^2)^{\frac{p}{2}}}\right)}\non
  &\ \ \leq 
m_+^2\frac{\calA_{p-2}(m_-,R)\calG{\left(\frac{\eta_- R}{2}\right)}+\calB_{p-2}(m_-,R)\calG{\left(\frac{3\eta_- R}{2}\right)}}{\min(\{c_n\})}\non
  &\phantom{=\ }
\ \   -m_+\eta_+R\frac{\calA_{p-1}(m_-,R)\calG'{\left(\frac{\eta_- R}{2}\right)}+\calB_{p-1}(m_-,R)\calG'{\left(\frac{3\eta_- R}{2}\right)}}{\min(\{c_n\})}\non
  &\phantom{=\ }
  \ \ +\frac{\eta_+^2R^2}{4}\frac{\calA_{p}(m_-,R)\calG''{\left(\frac{\eta_- R}{2}\right)}+\calB_{p}(m_-,R)\calG''{\left(\frac{3\eta_- R}{2}\right)}}{\min(\{c_n\})}\;,
  \end{align}
\endgroup
where $k_n\geq m_-+\eta_-n$ and $k_n\leq m_++\eta_+n$ and the
derivative with respect to the argument of $\calG$ is given by
Eq.~\eqref{eq:calGprime} and the double derivative yields 
\beq
\calG''(x) = \frac{e^{-x} + e^{-2x}}{(1 - e^{-x})^3}\;.
\eeq
\hfill$\square$
\begin{observation}
The infinite sum $\sum_{n=0}^{\infty} a_n$ with $a_n\geq 0$
remains finite by restricting to odd or even $n$.
\label{obs:1}
\end{observation}
\begin{corollary}
By means of the lemmata \ref{lemma:1}, \ref{lemma:2} and \ref{lemma:3}
and the observation \ref{obs:1}, all terms in the energies $E_{1,2}$
and $E_{3,4}$ of the cubic lattice in Eqs.~\eqref{eq:E12} and
\eqref{eq:E34}, are separately finite. 
\end{corollary}

\end{document}